\documentclass[a4paper,fleqn,usenatbib,twocolumn]{mnras}

\usepackage{newtxtext,newtxmath}

\usepackage[T1]{fontenc}
\usepackage{ae,aecompl}

\usepackage{graphicx}	
\usepackage{amsmath}	
\usepackage{amssymb}	

\newcommand{\HI}{\ion{H}{i}}

\title[21\,cm and CMB cross correlation]{Study of systematics effects on the Cross Power Spectrum of 21\,cm Line and Cosmic Microwave Background using Murchison Widefield Array Data}

\author[S. Yoshiura et al.]{
 S. Yoshiura$^1$\thanks{E-mail: 161d9002@st.kumamoto-u.ac.jp},
 K. Ichiki$^{2,3}$,
 B. Pindor$^{4,6}$, 
 K. Takahashi$^1$,
 H. Tashiro$^2$,
 C. M. Trott$^{5,6,7}$
\\
$^{1}$Department of Physics, Kumamoto University, Kumamoto,Japan\\
$^2$Department of Physics and Astrophysics, Nagoya University Furo-cho, Chikusa-ku, Nagoya, Aichi 464-8602, Japan\\
$^3$Kobayashi-Maskawa Institute for the Origin of Particles and the Universe, Nagoya University Nagoya, 464-8602, Japan\\
$^4$ARC Centre of Excellence for All-sky Astrophysics (CAASTRO)\\
$^5$International Centre for Radio Astronomy Research, Curtin University, Perth, WA 6845, Australia\\
$^6$School of Physics, The University of Melbourne, Parkville, VIC 3010, Australia\\
$^7$ARC Centre of Excellence for All Sky Astrophysics in 3 Dimensions (ASTRO 3D)
}

\begin{document}
\label{firstpage}
\pagerange{\pageref{firstpage}--\pageref{lastpage}}
\maketitle
\begin{abstract} 
Observation of the 21\,cm line signal from neutral hydrogen during the Epoch of Reionization is challenging due to extremely bright Galactic and extragalactic foregrounds and complicated instrumental calibration. 
A reasonable approach for mitigating these problems is the cross correlation with other observables. 
In this work, we present the first results of the cross power spectrum (CPS) between radio images observed by the Murchison Widefield Array and the cosmic microwave background (CMB), measured by the Planck experiment. 
 {We study the systematics due to the ionospheric activity, the dependence of CPS on group of pointings, and frequency.} 
The resulting CPS is consistent with zero because the error is dominated by the foregrounds in the 21\,cm observation. Additionally, the variance of the signal indicates the presence of {unexpected systematics} error at small scales. Furthermore, we reduce the error by one order of magnitude with application of a foreground removal using a polynomial fitting method. Based on the results, we find that the detection of the 21\,cm-CMB CPS with the MWA Phase I requires more than 99.95\% of the foreground signal removed, 2000 hours of deep observation and 50\% of the sky fraction coverage.
\end{abstract}

\section{Introduction}

At the end of the Dark Ages, the first stars were born in dense clouds and started to ionize neutral hydrogen ($\HI$). 
As a result of the cosmological structure formation, young galaxies and active galactic nuclei created large ionized bubbles around them. 
According to observations of high-$z$ QSO spectra, almost all of the $\HI$ gas was ionized by $z\,\sim \,6$ \citep{2006AJ....132..117F}. 
This is the epoch of reionization (EoR) and a measurement of its precise history is a primary motivation of current observational astrophysics.
There are other observations to measure the EoR. For example, the optical depth to the Thomson scattering of the Cosmic Microwave Background (CMB) photons constrains the duration of the reionization \citep{2016A&A...596A.108P}, and the decreasing of the Lyman-$\alpha$ emitter luminosity function yields the neutral fraction at $z>6$ \citep{2010ApJ...723..869O}. 

In particular, the 21\,cm line signal produced from the hyperfine structure of $\HI$ is a promising tool to probe the EoR. As the redshifted 21\,cm line is observed as a function of frequency, one can measure the 21\,cm line signal along the line of sight. Thus, the redshifted 21\,cm line provides the 3 dimensional distribution of $\HI$  gas, and we can study, for example, the nature of ionizing sources and precise history of reionization via the topology of ionized gases and its statistical property. The power spectrum is a common tool to characterize the parameters of the reionization model (e.g. \cite{2015MNRAS.449.4246G}). 

The redshifted 21\,cm line signal is observed by radio telescopes. There are many ongoing experiments, such as the the Giant Metrewave Radio Telescope EoR Experiment (GMRT, \cite{2013MNRAS.433..639P}), the Donald C. Backer Precision Array for Probing the Epoch of Reionization (PAPER, \cite{2010AJ....139.1468P}), the LOw Frequency ARray (LOFAR, \cite{2013A&A...556A...2V}), and the Murchison Widefield Array (MWA, \cite{2013PASA...30....7T,2013PASA...30...31B}). Meanwhile, the construction of future instruments such as the Hydrogen Epoch of Reionization Array (HERA, \cite{2016icea.confE...2D}) and the low frequency Square Kilometre Array (SKA\_{}LOW, \cite{2013ExA....36..235M}) is planned to begin in a few years. 
The current suite of interferometers have enough sensitivity for a statistical detection of the 21\,cm line signal. Although upper limits on the 21\,cm power spectrum at various scales and redshifts have been given by many groups \citep[e.g.][]{2015ApJ...809...61A, 2015PhRvD..91l3011D, 2014ApJ...788..106P, Jacobs2015,2016ApJ...833..102B,2017ApJ...838...65P}, the redshifted 21\,cm line signal has not been detected yet because of the insufficient sensitivity, complicated instrumental systematics and extremely bright foregrounds.

Typically, the foreground fluctuation is 3 $\sim$ 4 orders of magnitude larger than the 21\,cm line signal. To measure the 21\,cm line signal, one has to remove the synchrotron emission from our Galaxy and extragalactic point sources with high precision. As an useful character of the synchrotron radiation, the spectrum is expected to be smooth along the frequency axis. Thus, a classical method of foreground removal is the polynomial fitting along the line of sight. Although this method could be ineffective when the foregrounds have variance on the spectra, it is simple and pretty effective in the ideal case. 

In addition to the foreground removal, the cross correlation with other observable can mitigate the contamination from the foregrounds. 
For instance, the distribution of Lyman-$\alpha$ emitter correlates with the 21\,cm line signal \citep[e.g.][]{2009ApJ...690..252L,2017ApJ...836..176H} but not with the foregrounds. 
Thus, the foregrounds contribute to the observation as a source of error and can be statistically reduced. 
Although the foreground removal is required to mitigate the error \citep{2017arXiv170904168Y}, many works show that the MWA and the SKA1\_LOW have potential to detect the 21\,cm-LAE cross power spectrum \citep{Kubota}.

Furthermore, the cross correlation between the 21\,cm line signal and the CMB is also a useful observable. 
During the EoR, the peculiar motion of ionized bubbles generates the Doppler anisotropy to the scattered CMB photons. Because the 21\,cm line signal distribution reflects the topology of ionized bubbles, the Doppler anisotropy and the 21\,cm line signal have correlation at large scales ($l=100$). According to the analytic formulae, the expected 21\,cm-CMB signal consists of homogeneous part which is negative and in-homogeneous part which is positive. {Furthermore, the CMB includes other kSZ fluctuations such as the Ostriker-Vishniac effect \citep{1986ApJ...306L..51O} and patchy components (e.g. \cite{2016ApJ...824..118A}.) The angular power spectrum of the fluctuations has peaks at $l>300$, and the components can be a source of correlation with the 21\,cm signal. } There are many works predicting the signal and the expected signal amplitude is less than $10^4 \rm \mu K^2$ depending on, for example, the duration and the timing of reionization \citep[e.g.][]{2004PhRvD..70f3509C, 2006ApJ...647..840A,2005MNRAS.360.1063S,2008MNRAS.384..291A,2010MNRAS.402.2279J,2016ApJ...824..118A,2017arXiv171205305M}. Simultaneously with the signal prediction, the less detectability has been shown \citep{Tashiro2008}.

In this work, we calculate the 21\,cm-CMB cross power spectrum using MWA data and the CMB temperature map observed by the Planck experiment. \cite{Tashiro2008} have shown that a deep observation is required to detect the 21\,cm-CMB cross power spectrum using the MWA, as is such for the auto power spectrum. However, in order to reduce the noise effectively, systematic errors have to be less than the thermal noise. Then, to find unexpected systematics, an analysis using a common dataset is important before analyzing a large amount of data. Thus, the primary purposes of this work are to study the foreground and systematics contamination on the cross power spectrum analysis. Additionally, we attempt to remove the foregrounds from the MWA data to mitigate the systematic error. 


The structure of this paper is as follows. The MWA data and calibration are explained in Sec.~\ref{S:MWA}. In Sec.~\ref{S:method}, we introduce the 21\,cm-CMB cross power spectrum and calculation methods.  In Sec.~\ref{S:FGRM}, we describe our polynomial foreground removal method. In Sec.~\ref{S:system}, we investigate the contamination of ionosphere, the dependence on groups of pointings and on frequency.  The resulting auto and cross power spectra are shown in Sec.~\ref{S:results}. In Sec.~\ref{S:discuss}, we discuss the forecast. Finally, we summarize this work in Sec.~\ref{S:summary}.

\section{Murchison Widefield Array}\label{S:MWA}

The MWA is an interferometer constructed within the Murchison Radioastronomy Observatory in Western Australia. 
A tile consists of 16 dipole antennas and the effective area of antenna is about 20$\rm m^2$ at 150MHz. 
The observable frequency range is 80MHz to 300MHz, and instantaneous bandwidth is 30.72MHz which is divided by 1.28MHz coarse channels. 
The maximum baseline of the MWA phase 1 is about 3000m and the maximum resolution is a few arcminutes. 
The MWA has huge primary beam (400$\rm deg^2$) and large side lobes. The large field of view (FoV) is an advantage of the MWA for the cross correlation because the large FoV can reduce the cross correlation error. Further details are written in \cite{2013PASA...30....7T}. Note that, until middle of 2016, 128 tiles operated as one interferometer so-called the MWA Phase I. The MWA Phase II observation is ongoing with additional 128 tiles and different configuration of tiles. 

In this work, we use the radio data observed by the MWA Phase I in 2013. The pointing center is toward the EoR0 field centered at (RA,Dec) = (0h, $-$27deg), which was chosen to be sufficiently far from the Galactic Center. The observed frequency range is 167MHz to 197MHz and frequency channel is 40kHz. 
The observation is operated around zenith for 3 hours and the data consists of 90 observations with 112s time interval for each of snapshots. 
During the ``drift and shift'' observation, the direction of antenna pointing changes 5 times, with a change every 30 minutes toward the center of the EoR0. It is worth noting that the beam shape of the MWA varies with the pointing \citep{2016ApJ...833..102B,2016ApJ...819....8P}.  

Although radio frequency interference (RFI) in the Murchison Radioastronomy Observatory is minimal, 1.1 \% of data are polluted \citep{2015PASA...32....8O}. Using the AOFlagger \citep{2010MNRAS.405..155O}, the data including the RFI are flagged and removed from each snapshot. For calibrating the data, we use the Real Time System (RTS, \citep{Mitchell2008,Ord2010}), which is a primary calibration software pipeline for the MWA. There are many articles explaining the detail \cite{Mitchell2008,Jacobs2016}. We describe the brief method of the RTS below. 

The RTS calibration technique is called ``Peeling method'', based on the catalogue of apparent bright sources in southern sky and performed in visibility space. The point source catalogue of the EoR0 field is build using the the Positional Update and Matching Algorithm (PUMA) \citep{2017PASA...34....3L}. The catalogue is used in the main calibration process of the RTS, named Calibrator Measurement Loop (CML). During the CML process, firstly, the models visibilities of all sources are subtracted from the data. Secondly, the visibilities of the target sources are returned into the subtracted data. Then, the data is dominated by the target source, and one can measure the ionospheric offsets from a phase ramp by fitting to the visibility phased to the position of the aimed source. In this work, the ionospheric offset is corrected using 1000 point sources. Finally, 5 brightest sources in the catalogue are used to correct the direction-dependent antenna gains. After the calibration, the 1000 point sources are peeled from the visibility.  Using the calibrated visibilities, the RTS imager provides the snapshot images with wide-field corrections.

\begin{figure*}
\centering
\includegraphics[width=7cm]{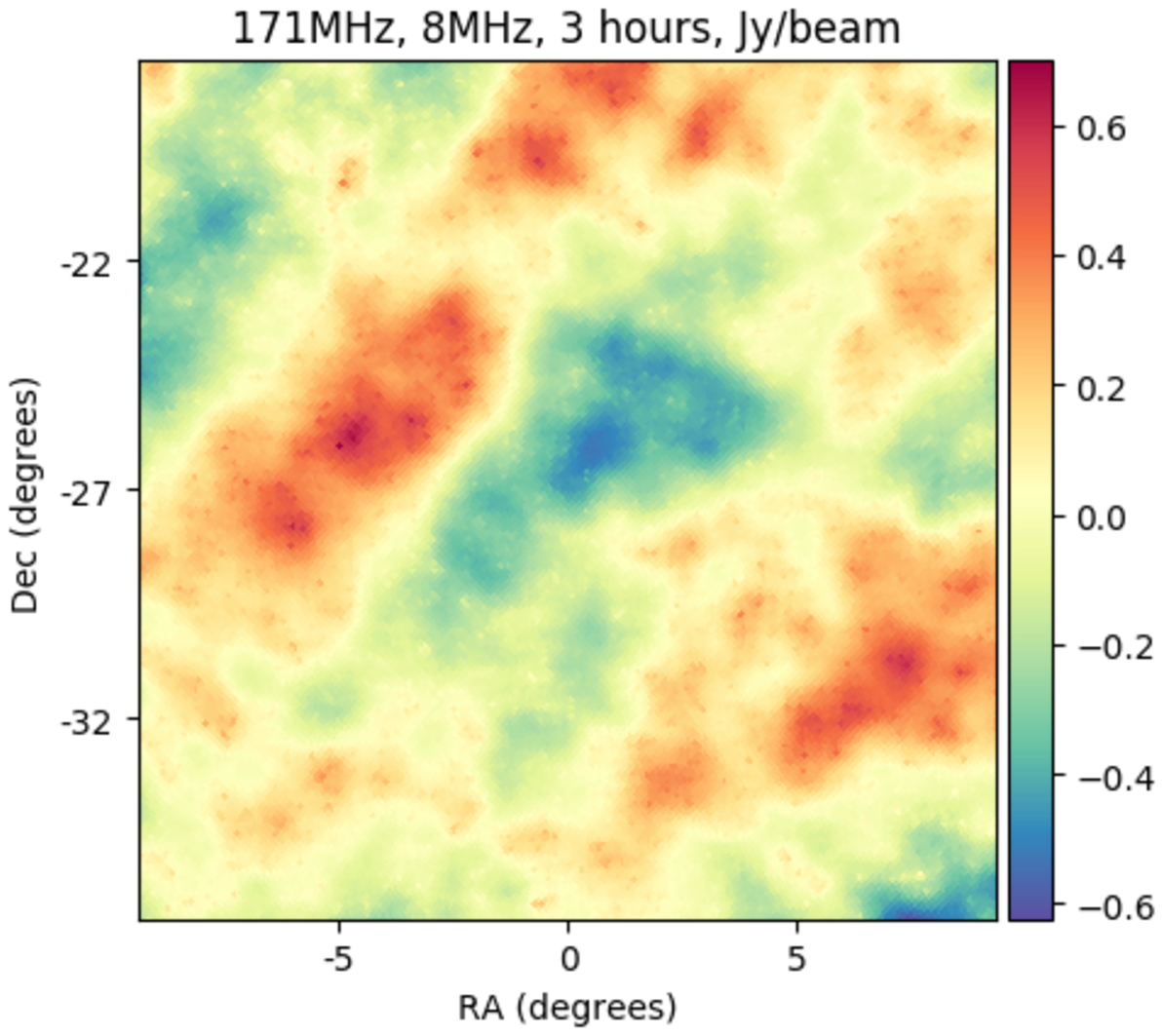}
\includegraphics[width=7.5cm]{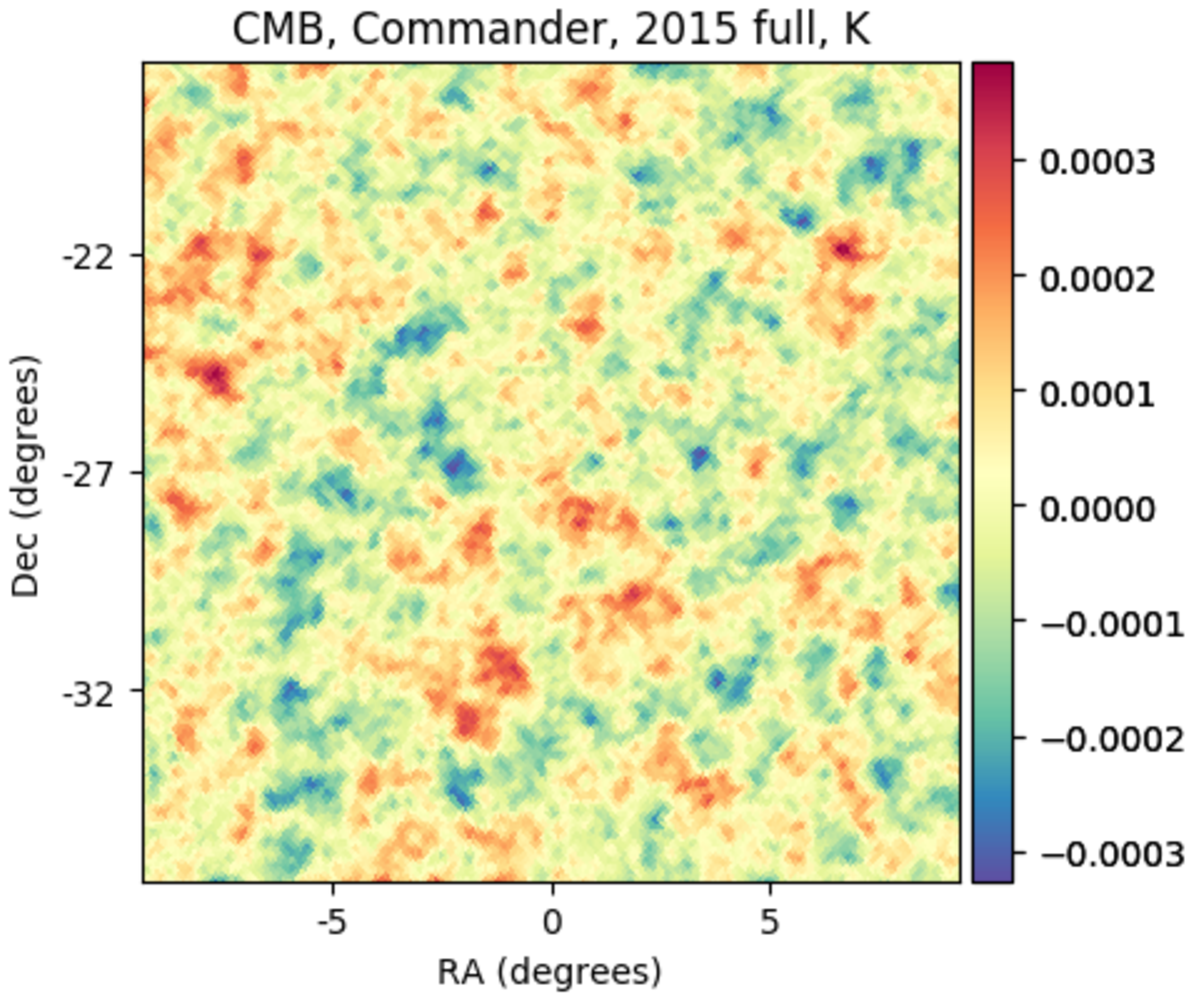}
\caption{Left panel is the observed image centered at EoR0 field, (RA, Dec) = (0, $-$27), with $\Delta B\,=\, \rm8\,MHz$, $t_{\rm int}\,=\, \rm 3\,hours$. The field of view is $\sim (20\,\rm deg)^2$ and the centered frequency is 171$\,$MHz. This image is made from 540 snapshots created by RTS with natural weights. Using the RTS, 1000 bright point sources are peeled. The image shows apparent diffuse structure and un-subtracted point sources. Negative value is a result of lack of sampling at the center of uv-plane. Right panel is the CMB temperature fluctuation map corresponding to the EoR0 field.} 
\label{fig:image} 
\end{figure*}

We obtain output images at every snapshot (112\,s) and coarse band channel (1.28\,MHz) in the HEALpix frame \citep{2005ApJ...622..759G}. For reducing the thermal noise contamination, 90 observations are integrated into a main image at every $\sim$ 8MHz, namely a main image includes 540 snapshots. Left panel of Fig.~\ref{fig:image} is the image centered at 171MHz using natural weighting. This image is almost consistent with the Figure.~8 in \cite{2016ApJ...833..102B} except the size and resolution. There is apparent diffuse emission and un-subtracted point sources. Note that the lack of visibility at $(u,v) = (0,0)$ (no auto-correlations) causes the negative values, whereby a constant sky temperature cannot be reconstructed. 

In addition to the images, the calibrated visibility data are an output of the RTS. In this work, we take the average of the visibilities in the uv plane with $\Delta u = 1$ at every coarse band channel. These gridded visibilities are used to calculate the angular power spectrum.

\section{Methods}\label{S:method}

In this section, we describe the method to calculate the angular power spectrum (APS), cross power spectrum (CPS) and the error of the CPS. 

\subsection{Angular Power Spectrum}
The APS is defined as ensemble average of spherical harmonics expansion coefficients,
\begin{eqnarray}
\langle a_{lm}a_{l'm'}^* \rangle = \delta_{ll'}\delta_{mm'} \langle C_l \rangle,
\end{eqnarray}
where $\delta_{ii'}$ is Kronecker symbol and,
\begin{eqnarray}
a_{lm}(\nu) = \int  d{\Omega} ~ Y_{lm}^* ~ I (\nu,{\bf{\theta}}).
\end{eqnarray}
Here $ I (\nu,{\bf{\theta}})$ is intensity of the images.

To calculate the APS from images, we use Polspice\footnote{http://www2.iap.fr/users/hivon/software/PolSpice/}\citep{Chon2004}, which is originally developed for estimating the CMB polarization power spectrum. The APS is evaluated via weighted correlation function of images, and full method is described in \cite{Chon2004}. The Polspice has been used in some literatures such as the BICEP 2yr data analysis \citep{2010ApJ...711.1123C}, the CMBPol mission concept study \citep{2009AIPC.1141..222D} and the angular power spectra of the Planck frequency maps \citep{2016A&A...594A..11P}. 
Note that an image created from visibility has weight effect caused from the uv-coverage. 
For correct estimation of APS, we have to correct the weight. The correction method is described in the Appendix.

The units of the images made by using the RTS is Jy/beam, and we need to convert Jy/beam into K. However, the conversion is not straightforward because it needs an accurate beam model. Thus, in this work, we normalize the APS of the images using a constant factor so that the image APS is consistent with the APS calculated from the visibilities at $\ell = 500$. 

The APS of visibility data is computed as \citep{2012ApJ...757..101T}, 
\begin{eqnarray}
C_{l} = \frac{\displaystyle{\sum_{2\pi|u|=l}}N_{uv}|V_{uv}|^2}{\displaystyle{\sum_{2\pi|u|=l}}N_{uv}},
\label{Clvis}
\end{eqnarray}
where $N_{uv}$ is the number of visibility samples at a uv-grid and $V_{uv}$ is the averaged visibility.
The units of visibility is Jy, and we convert Jy into K as,
\begin{eqnarray}
{\rm K} = \frac{\lambda^2}{2\times10^{26}k_{\rm B}\Omega}[\rm Jy]
\label{Jy2K}
\end{eqnarray}
where $\lambda$ is observed wavelength, $k_{\rm B}$ is the Boltzmann constant and $\Omega$ is the solid angle of the MWA's field-of-view. 

The APS using the visibility data can be used to estimate the error power of the 21\,cm observation. For the error estimation, we split the data into ``even'' and ``odd'' samplings at 8s time intervals, and calculate the APS of the difference between the samplings. Ideally, the 21\,cm signal and the foreground radiation are common in those two groups while the thermal noise is time variant. Therefore, the difference should be dominated by the thermal noise.

\subsection{Cross Power Spectrum}

Using the Polspice, we calculate the CPS from images of the MWA radio map and the CMB temperature fluctuation map. The definition is written as
\begin{eqnarray}
\langle a^{21}_{lm}a^{\rm CMB,*}_{l'm'} \rangle = \delta_{ll'}\delta_{mm'} \langle C^{21,\rm CMB}_l \rangle,
\end{eqnarray}
where $a^{21}_{lm}$ and $a^{\rm CMB}_{lm}$ are calculated from the MWA image and the CMB image. 
The images are cut into square of 20$\times$20 $\rm deg^2$ and the masked field could be a unique response on the signal. The mask effect is reduced using the Polspice \citep{Chon2004}. Same as the APS, we correct the weight caused from incomplete uv-coverage. 

The CPS should be computed from the visibility to account for uv-coverage and avoid the systematics caused from imaging. For the calculation, one has to Fourier Tranform the CMB image into the visibility uv-plane. In this work, the CPS is calculated from images because we remove foreground using the line of sight fitting method. Computation via visibilities is left for future work.

The error of CPS is estimated using \citep{1995PhRvD..52.4307K},
\begin{eqnarray}
\label{eq:error1}
\Delta C^2_{l} &=& A\left[\left(C_{l}^{\rm 21,CMB}\right)^2+\left(C_{l}^{\rm 21,obs}\right)\left(C_{l}^{\rm CMB,obs}\right)\right],\\
&=&A\left[\left(C_{l}^{\rm 21,CMB}\right)^2+\left(C_{l}^{\rm 21}+N_{l}^{\rm 21}+C_{l}^{\rm FG}\right)\left(C_{l}^{\rm CMB}+N_{l}^{\rm CMB}\right)\right],\nonumber
\end{eqnarray}
where $A^{-1} = (2l+1)f_{\rm sky}\Delta l$, $f_{\rm sky}$ is the sky fraction of the image, $\Delta l$ is the bin width, $C_{l}^{\rm MWA,obs}$ is the APS of the images observed by the MWA and $C_{l}^{\rm CMB,obs}$ is the APS of the CMB image. The error consists of a sample variance term and a contribution of 21\,cm APS, noise of the 21\,cm observation, foreground APS, the CMB APS and the noise of the CMB observation. In this work, we ignore the error term of the sample variance, which should be smaller than other terms. In practice, the CMB temperature map includes the foreground residual, which can correlate with the foreground of the 21\,cm observation. However, the foreground contamination is sufficiently smaller than the CMB signal, and therefore we ignore this term. In addition, we consider the error propagation as described in the Appendix because we use the weighted CPS. 

For the calculation, we use the 2015 CMB full mission Commander map stored in Planck Legacy Archive \footnote{Based on observations obtained with Planck (http://www.esa.int/Planck), an ESA science mission with instruments and contributions directly funded by ESA Member States, NASA, and Canada.}. The right panel of Fig.~\ref{fig:image} shows the CMB map associated with the EoR0 field.

\section{Foreground removal}\label{S:FGRM}

As shown in Eq. \ref{eq:error1}, the foregrounds contribute to 21\,cm-CMB cross power spectrum as a source of error. 
Since the foregrounds are a few orders of magnitude brighter than other contributions, the terms of the foreground and CMB PS dominate the total error. Thus, one has to remove the foreground from the MWA images. 
Although 1000 bright point sources have been subtracted from the MWA data, the diffuse emission and un-peeled sources remain in the data. 
In this work, we perform a simple polynomial fitting to the spectral dimension of the data, and use this as the foreground removal.

First, we make 24 images with 1.28~MHz spectral channels from the MWA data in the observed frequency range of 167~MHz to 197~MHz. Next, we fit a $n$-th-order polynomial function along the frequency axis at each pixel of the image cube, and obtain smooth functions. Finally, we subtract the smooth functions from our main images.

\begin{figure*}
\centering
\includegraphics[width=8.5cm]{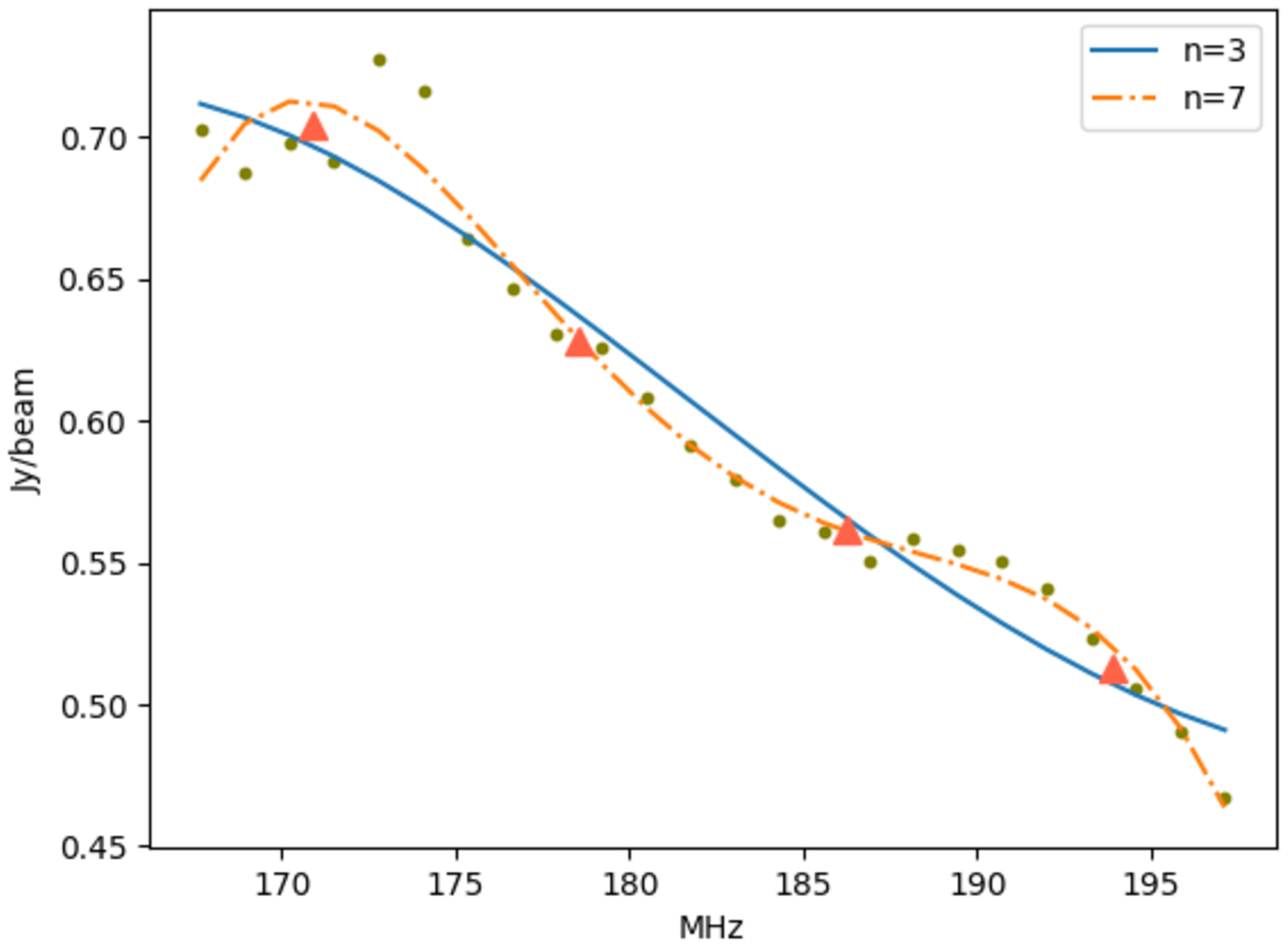}
\includegraphics[width=8.5cm]{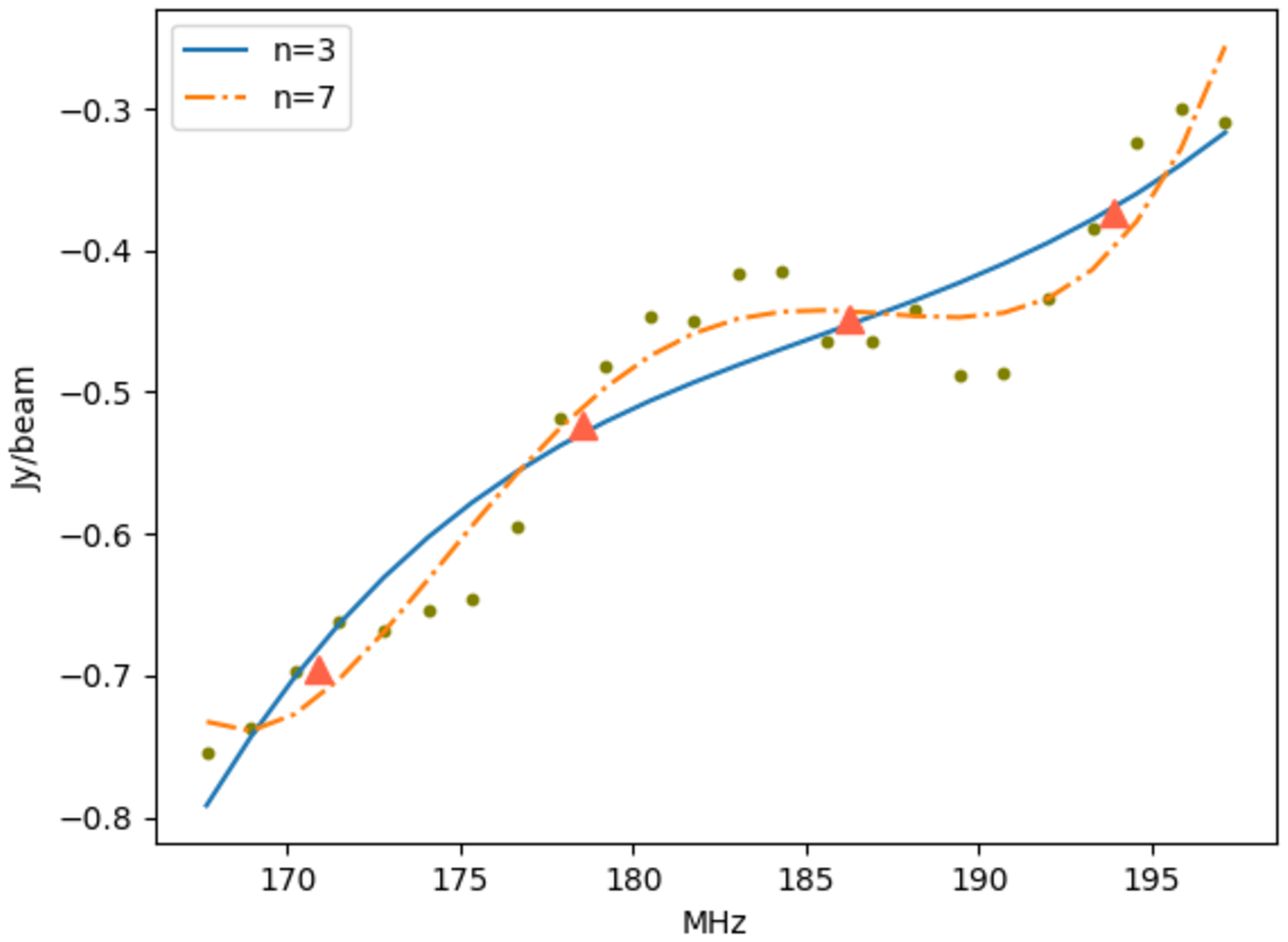}
\caption{Examples of foreground fitting along the line of sight. Left and right panels show the case of the brightest pixel and the faintest pixel at $\nu=167$MHz. The green circles are values of the images with 1.28~MHz as $\Delta B$, and the orange triangles are values of the pixels from images with 8~MHz bandwidth. The solid, dashed and dot-dashed lines are obtained by fitting to 24 circles with $n$=3 and 7.} 
\label{fig:fit} 
\end{figure*}

The left (right) panel of Fig.~\ref{fig:fit} shows examples of the fitting for the brightest (faintest) pixel at 167~MHz. Circles are values at a pixel of each image for the fitting, and triangles are values of main images. Clearly, the circles are not smooth and have a wave-like feature as a function of frequency.  This feature is caused from the mode-mixing and frequency dependence of the instrumental point spread function (e.g. \cite{2012ApJ...752..137M}). In the images transformed from visibility data, there is a concentric side lobe structure around point sources because of the sparse uv-coverage, and the side lobe propagates emission from point sources in the field. Perfect removal of this structure is impossible due to the number of point sources. 

In the panels, the fitting function with $n$=3 does not reproduce the data and a higher $n$th-order polynomial function is required \footnote{Since the diffuse emission should be spectrally smooth, the polynomial fitting method is effective. However, to remove the point source contribution which is the wave-like spectrum, we may need additional functions. }. Thus, the remaining signal could be corrupted by the foreground residuals. 
In order to reduce the residual contamination, we attempt to fit with $n=7$ and the reproduction of the data is better than $n=3$. However, the $7$-th polynomial fitting slightly improves the results discussed following section. {Thus, we adopt $n=7$ for the foreground removal in this work.} Note that the higher order polynomial fitting function could fit out the EoR signal. 

Generally, the expected spectral index of the synchrotron emission is negative. However, as shown in the right panel of Fig.~\ref{fig:fit}, the data shows a positive index, which is understood as below. The average intensity is subtracted from actual intensity since the interferometer cannot observe the visibility at $(u,v) = (0,0)$. Then, for example, the intensity of image is {$F (\nu, \,\theta)-F_{\rm ave}(\nu)$}, where true flux of a LOS is {$F(\nu, \,\theta)=F_0\nu^{\alpha}$} and averaged flux is $F_{\rm ave}(\nu) = F_{\rm ave,0} \nu^{\overline{\alpha}}$. If the index $\alpha$ is larger than the $\overline{\alpha}$, the index of observed intensity can be positive. 

\begin{figure}
\centering
\includegraphics[width=7cm]{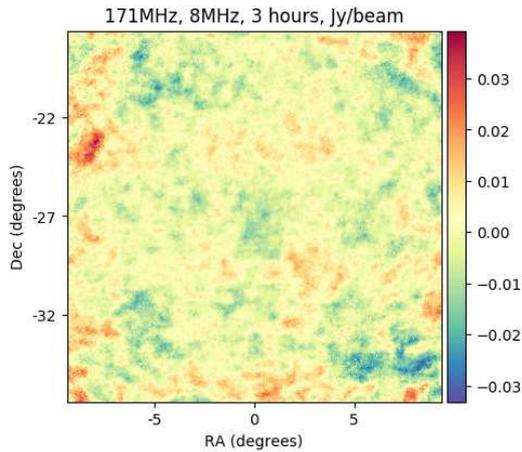}
\caption{The EoR0 field at 171~MHz with $\Delta B=8 \,\rm MHz$, $t_{\rm int}= 3 \,\rm hours$, $(20\,\rm deg)^2$. We remove the foregrounds by the polynomial fitting with $n=7$ to the image shown in Fig.~\ref{fig:image}. The large scale features have been removed and the fluctuation is reduced by one order of magnitude. } 
\label{fig:FGRM} 
\end{figure}

Fig.~\ref{fig:FGRM} shows the image at 171~MHz after performing the polynomial foreground removal with $n=3$. Interestingly, although the foreground removal method is relatively simple and has no prior that uses spatial correlation, we can reduce the diffuse structure shown in the left panel of Fig.~\ref{fig:image} and reduce the fluctuation by one order of magnitude. 

\begin{figure}
\centering
\includegraphics[width=7cm]{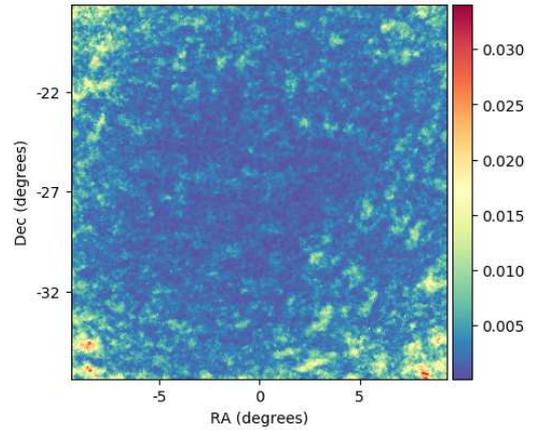}
\caption{{The squared error of the polynomial foreground removal with $n=7$. In the center of EoR0 field, the polynomial fitting works properly. }} 
\label{fig:chi2} 
\end{figure}

The quality of fitting depends on regions because the smoothness is different for each LOS. 
In this work, we quantify the quality of the fitting using squared error, $\sum_i (d_i-f_i)^2$, where $i$ denotes the number of frequency bins, $d_i$ denotes  the data and $f_i$ is the fitting function. Fig.~\ref{fig:chi2} shows the distribution of the error, and the spectral fitting works well in the center of the region. 
Thus, we calculate the APS using a region centered at (RA, Dec) = (-0.133 h, -28 deg) with a 4~degree radius, and then the resulting APS is reduced by 2 times. However we do not use the mask for simplicity and reducing the error of CPS which is proportional to $f_{\rm sky}^{-0.5}$. 

\section{Addressing Systematics}\label{S:system}

{
21\,cm observations with interferometers suffer from systematics such as refraction due to electrons in the Earth's ionosphere, instrumental systematics, and intrinsic foreground spectra. Thus, a better understanding about the systematics of the data is worthwhile for future further analysis. 
In this section, we study the effect of ionosphere, pointing of observation and frequency dependence of the CPS by comparing with that of the APS. 
}

\subsection{Pointing Dependence}

\begin{figure*}
\centering
\hspace{0cm}\includegraphics[width=7cm]{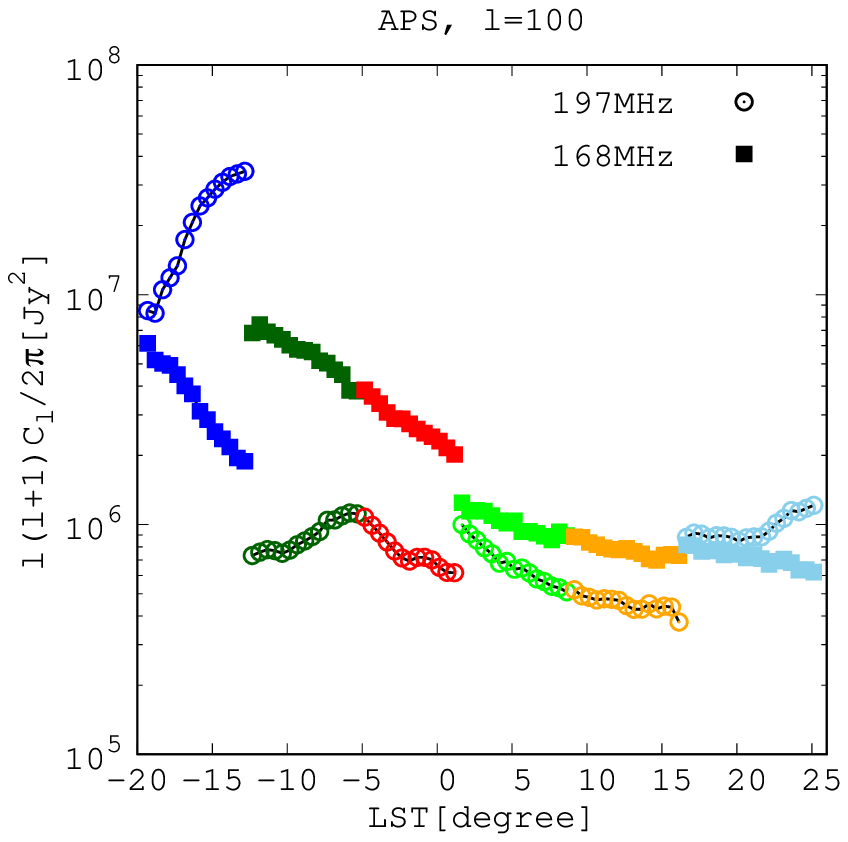}
\hspace{1cm}\includegraphics[width=7cm]{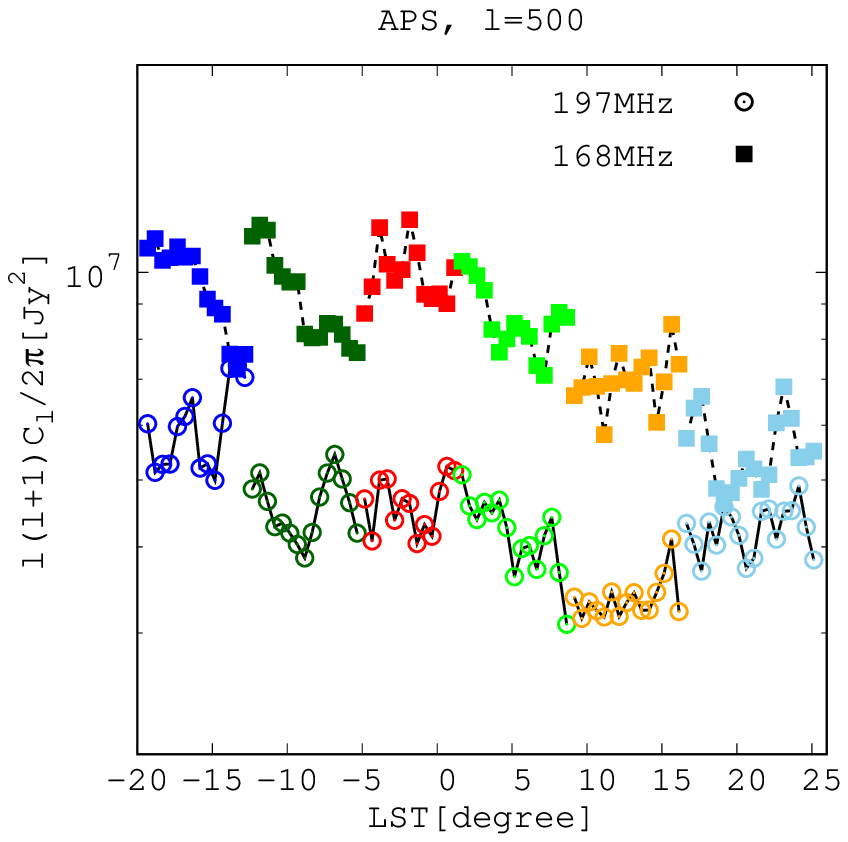}
\caption{ The LST dependence of APS at $l=100$ and $l=500$. Each points are the APS of visibility data with $\Delta B=1.28\,\rm MHz$ and $t_{\rm int}=$112s. Colors indicate different group of pointing. Because the spectral index of foreground is negative, the APSs of higher frequencies are smaller than that of lower frequencies. However, the APS of 197~MHz exceeds that of 168~MHz at large scales. This is due to the Galactic emission, which leaks to the data at the lowest LST because the MWA side lobe moves toward Galactic center and have large gain at high frequency.  } 
\label{fig:APSLST} 
\end{figure*}

The MWA observation is 3 hour continuous observation around zenith, and the signal can be calculated as a function of the Local Sidereal Time (LST). Furthermore, the observation strategy is the drift and shift observation, and the direction of pointing changes 5 times toward the center of the EoR0 field. Fig.~\ref{fig:APSLST} shows the LST dependence of the APS calculated from 90 visibility data sets. The colors represent the group of pointing. The right panel shows that the APS at 168~MHz is larger than the APS at 197~MHz on the small scale. On the other hand, as shown in the left panel, the APS at 197~MHz exceeds the APS at 168~MHz in range of $\rm LST< -13$ and $\rm LST > 16$ on large scale. During observation with these LST, the pointing is off-zenith and then the MWA has large side lobes (e.g. \citep{2015RaSc...50..614N}). Furthermore the gain of the side lobe is high at high frequency, and the second side lobe is close to the Galactic center. Thus, the enhancement of the APS at 197~MHz on large scale shows the leakage of the Galactic synchrotron emission. 

\begin{figure*}
\centering
\hspace{0cm}\includegraphics[width=7cm]{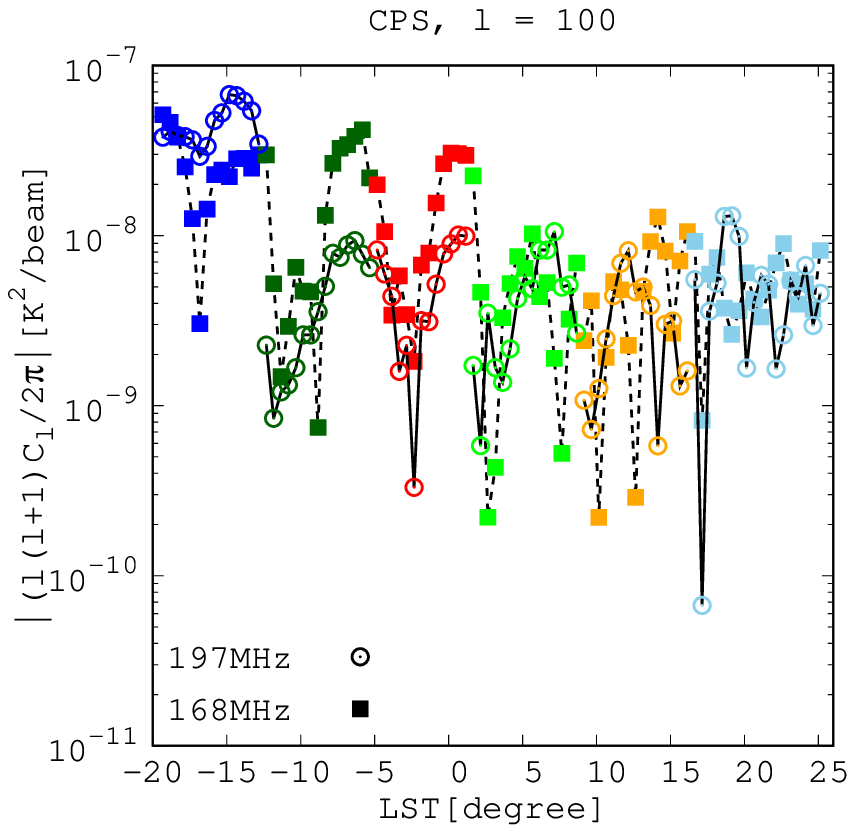}
\hspace{1cm}\includegraphics[width=7cm]{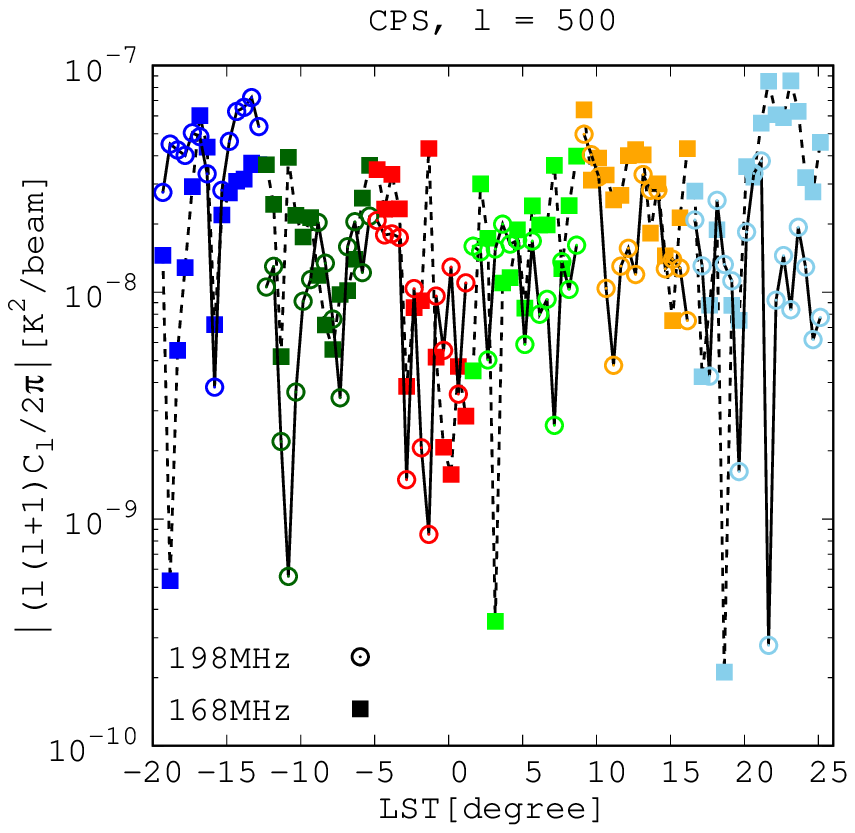}
\hspace{0cm}\includegraphics[width=7cm]{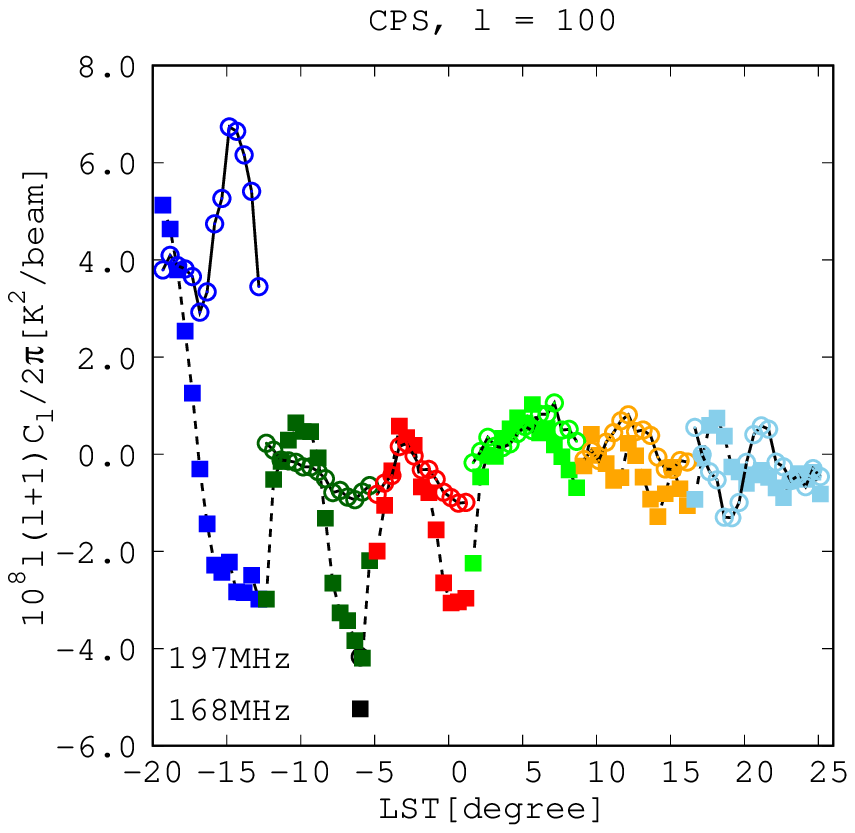}
\hspace{1cm}\includegraphics[width=7cm]{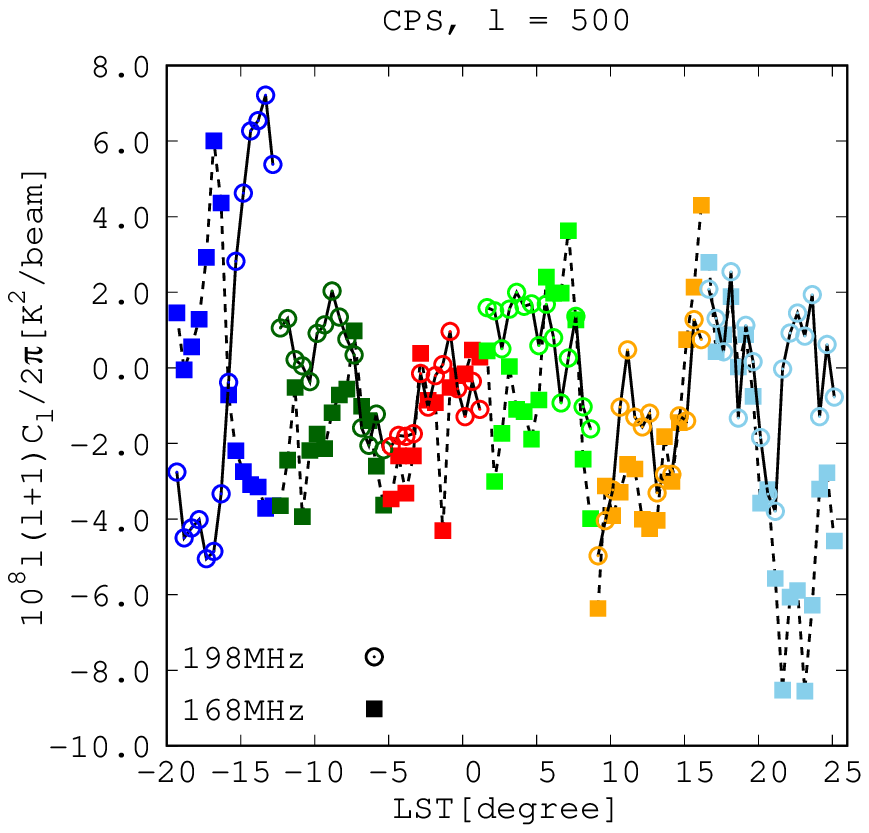}
\caption{{The LST dependence of CPS at $l=100$ and $l=500$. Each points are the CPS of image data with $\Delta B=1.28\,\rm MHz$ and $t_{\rm int}=$112s. Same as Fig.~\ref{fig:APSLST}, Colors indicate different group of pointing. Top panels show the absolute value of CPS, and bottom panels show the CPS in linear scale.} }
\label{fig:CPSLST} 
\end{figure*}

{
The top left panel of Fig.~\ref{fig:CPSLST} shows the LST dependence of the absolute CPS at $l=100$.  As is the case for the APS, the CPS increases at LST $<$ --13 degrees where the APS has serious foreground contamination, and the CPS at 197~MHz is larger than the CPS at 168~MHz. The LST dependence is apparent in the bottom left panel, which shows the CPS in linear scale. Apart from foreground contamination at LST $<$ -13 degrees, the amplitude of CPS at 167~MHz is larger than the CPS at 197~MHz, and CPS oscillates around zero. These results indicate that the CPS is dominated by errors due to foregrounds, which should be proportional to $(C_l^{\rm FG})^{1/2}$. Same as left panels, blue circles indicate the foreground contamination on the CPS at $l=500$ in the bottom right panel of Fig.~\ref{fig:CPSLST}. 
}

{
}


\subsection{Ionosphere Dependence}

{
The Earth's ionosphere refracts radio waves and produces phase errors and spatial offsets of radio point sources in interferometer measurements. Although the RTS calibrates the phase error, the uncalibrated error can produce residuals of point sources after the point source subtraction. 
Here, we show the effect of ionosphere activity on the APS and CPS. 
}

{
In \cite{Cris2017}, they have discussed particular types of ionosphere activity by defining the metric of ionosphere quality, $m$, based on the total offsets of point sources and dominant eigenvalue determined by a principal component analysis to the source offsets. In general, the quiet ionosphere has $m<4$ and the active has $m>10$. { Since, for example, the calibration error due to the ionosphere can generate the error of the point sources subtraction, then we might find a correlation of the metric value with the APS and the CPS. The metric is measured every 2 min observation, and} the metric of 63 observations, used in this work, is known.
}

\begin{figure*}
\begin{center}
\includegraphics[width=7cm]{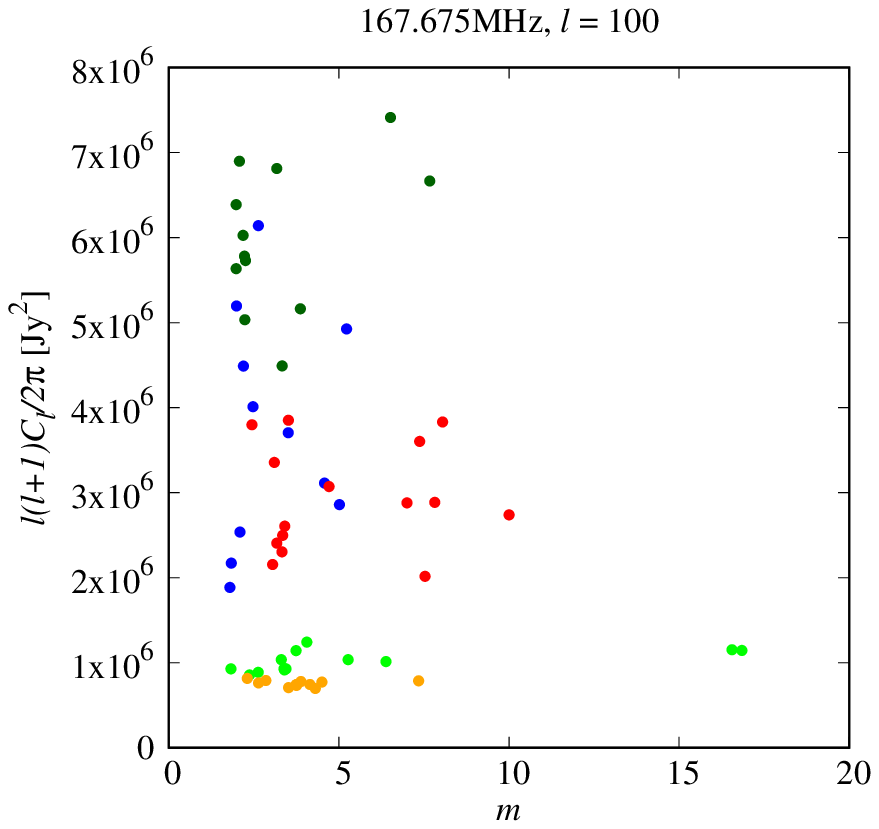}
\includegraphics[width=7cm]{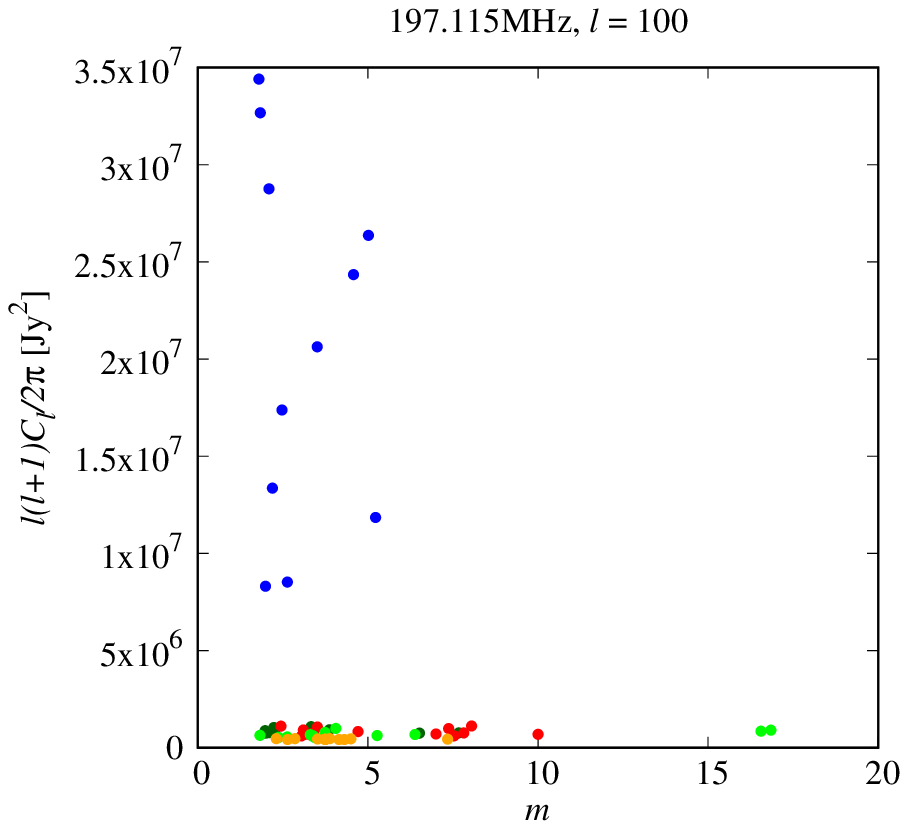}
\includegraphics[width=7cm]{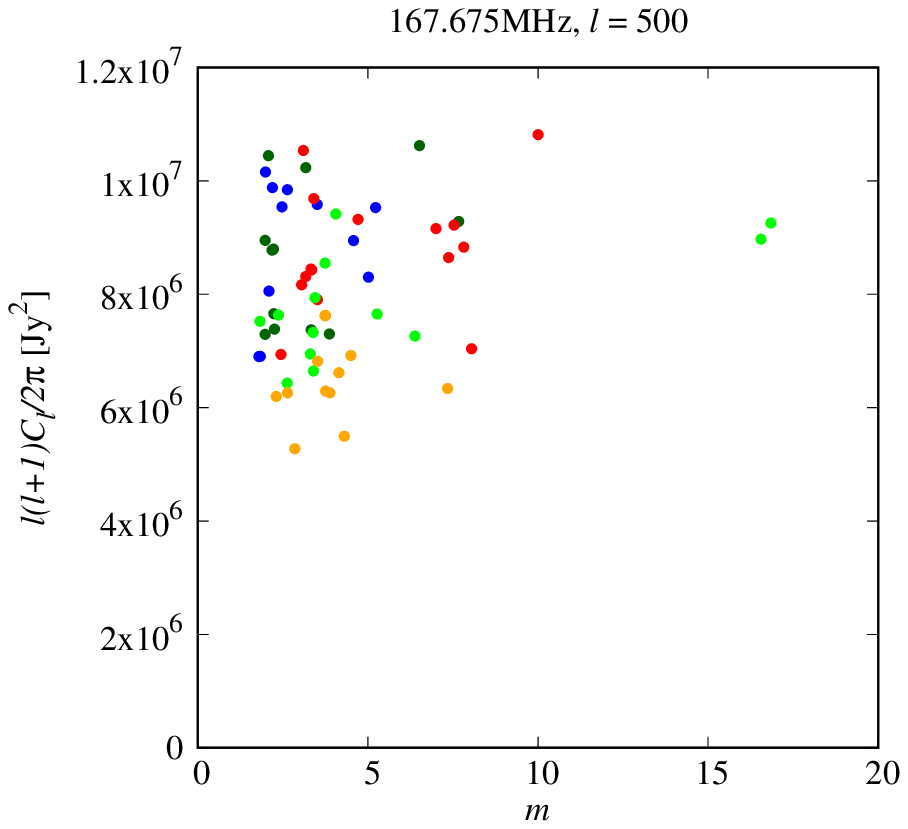}
\includegraphics[width=7cm]{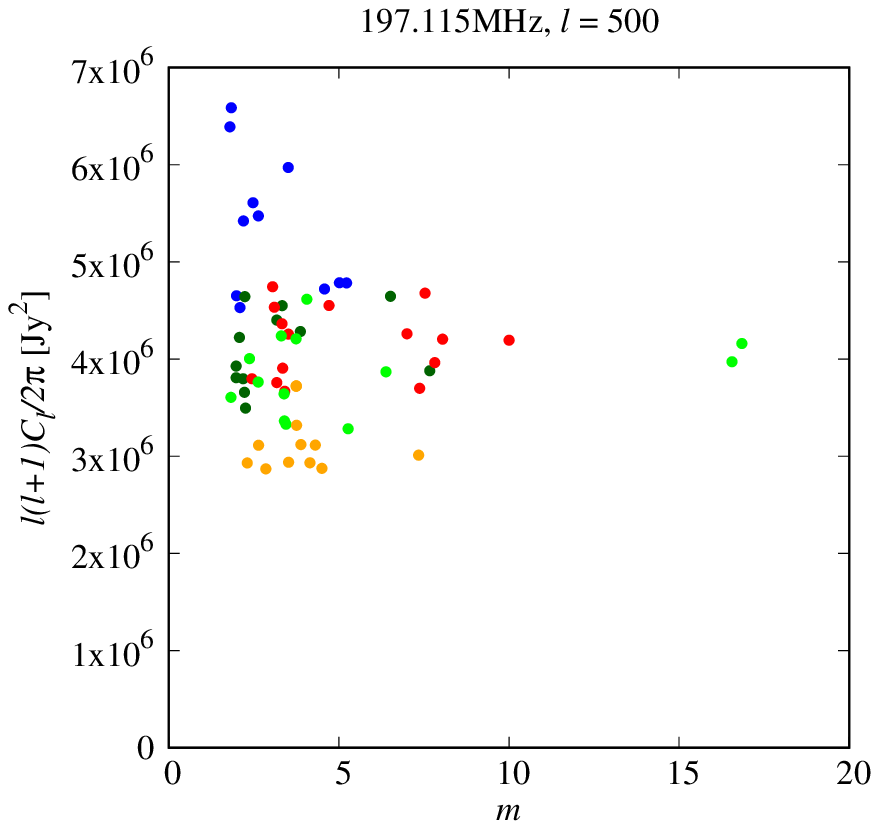}
\end{center}
\caption{{Ionosphere activity dependence of APS at $l=$100 (top) and 500 (bottom). We show the APS at 167.675~MHz and 197.115~MHz in left and right panels.  Each dots are the APS of image data with $\Delta B$=1.28 MHz and $t_{\rm int}$=112s. Same as Fig.~\ref{fig:APSLST}, different color shows different group of pointings.}}
\label{fig:IonoA} 
\end{figure*}

{
In Fig.~\ref{fig:IonoA}, we show the ionosphere metric dependence of the APS at 167.675 MHz and 197.115 MHz, and at $l$=100 and 500. The colour of dots represents pointing of observation as with the Fig.~\ref{fig:APSLST}. The metric of ionosphere does not have apparent correlation even with active ionosphere, $m>10$. Furthermore, APS with quiet ionosphere, $m<4$, and moderate, $4<m<10$, do not have particular dependence on $m$. This indicates that contamination due to ionosphere is smaller than the foreground residuals such as fainter point sources and diffuse foregrounds. On the other hand, pointing is a crucial systematic on the APS. As shown in right panels, the blue dots are stronger than others which is also indicated from Fig.~\ref{fig:APSLST}. 
}

\begin{figure*}
\begin{center}
\includegraphics[width=7cm]{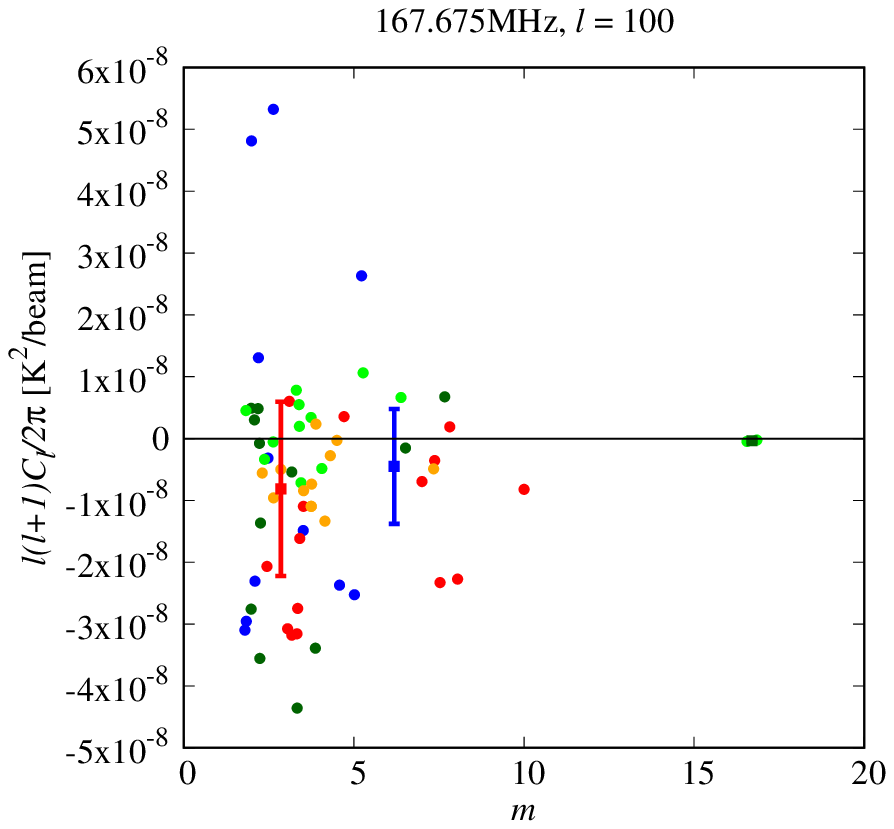}
\includegraphics[width=7cm]{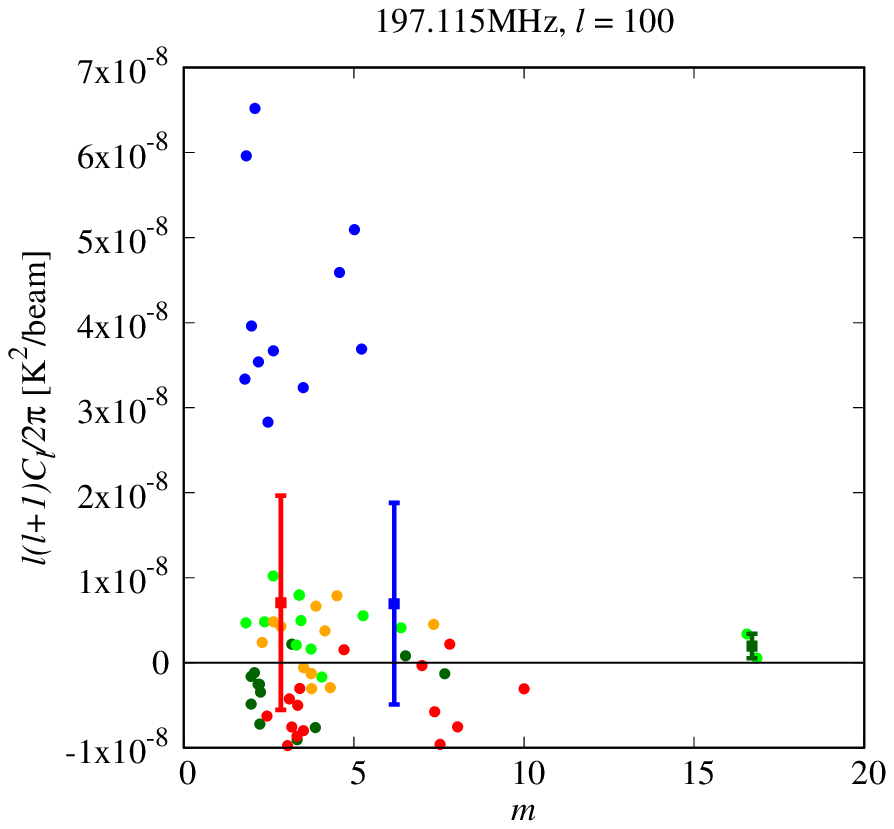}
\includegraphics[width=7cm]{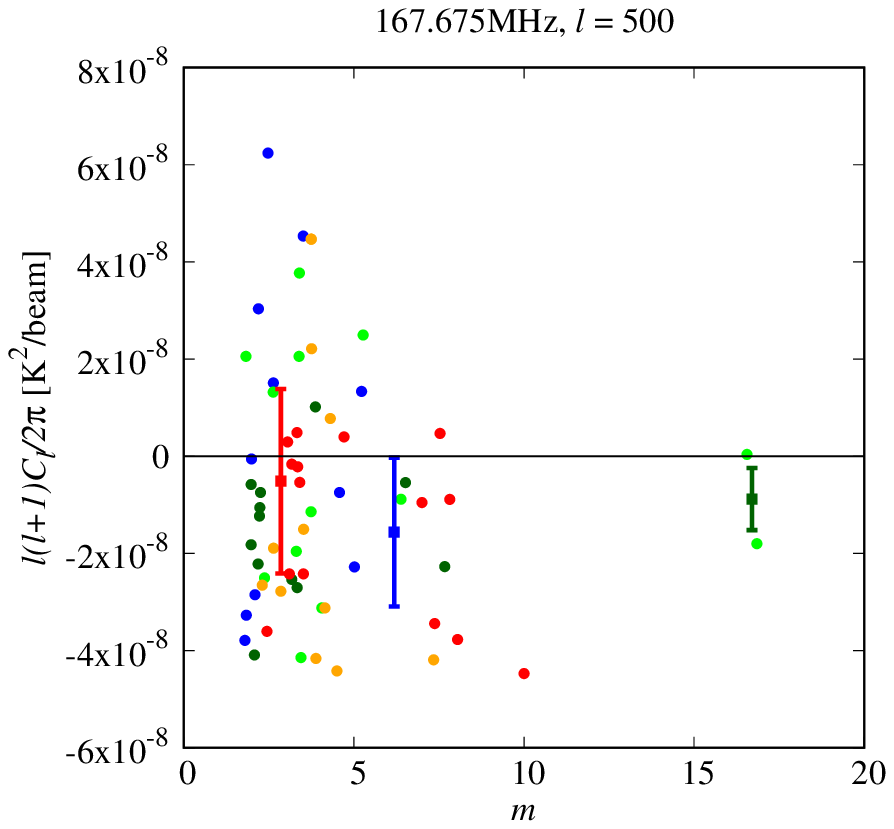}
\includegraphics[width=7cm]{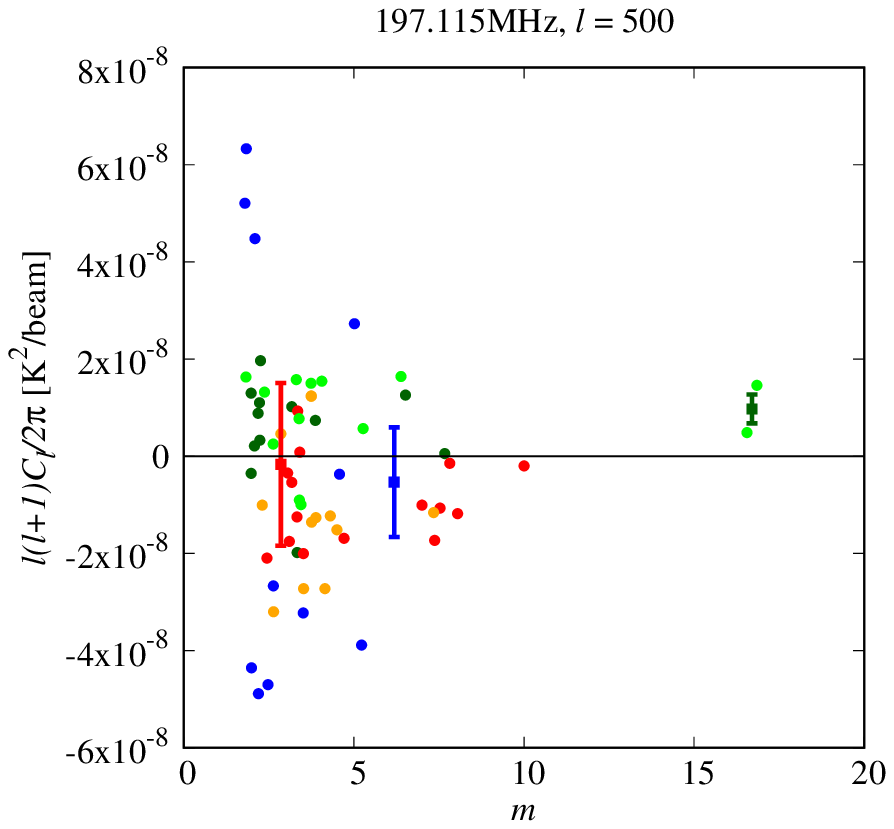}
\end{center}
\caption{{Same as Fig.~ \ref{fig:IonoX}, but for the CPS. 
Red, blue and green squares and error bars show average value and standard deviation of CPS with quiet, moderate and active ionospheres, respectively. }}
\label{fig:IonoX} 
\end{figure*}

{
In addition, the ionosphere effect on the CPS is shown in Fig.~\ref{fig:IonoX}, and also there is no apparent correlation between CPS and $m$. In the plot, we show the averaged CPS of quiet, moderate and active ionospheres with the standard deviation as error bars. The averaged CPS with quiet and moderate ionospheres are fairly consistent with zero, but the CPS with active ionosphere is not. Thus, the ionosphere might affect the CPS, but there are only a few samples observed in the same group of pointings. Furthermore, same as Fig.~\ref{fig:IonoA}, the dependency on sets of pointings can be found in the top right panel of Fig.~\ref{fig:IonoX}. This shows the contamination of leaked foreground on the error of CPS. 
Before moving to the next analysis, it is worth noting that the CPS oscillates around zero in Fig.~\ref{fig:IonoX}. This indicates that the CPS is dominated by the error including thermal noise and foreground.
}

{Systematics due to the ionosphere should be more important at smaller scales. However, we also do not find any correlation at $l$=1000 (not shown in this work) just as for $l$=100 and $l$=500. We note that only two of our samples have an active ionosphere, and an analysis with a more representative sample of ionospheric conditions will be required to more fully investigate these effects. We leave this for future work. }


\subsection{Frequency Dependence}

{The APS and CPS are expected to be dominated by the foregrounds, which depend on frequency. Here, we show the frequency dependence of the MWA data.}

\begin{figure}
\centering
\includegraphics[width=7.5cm]{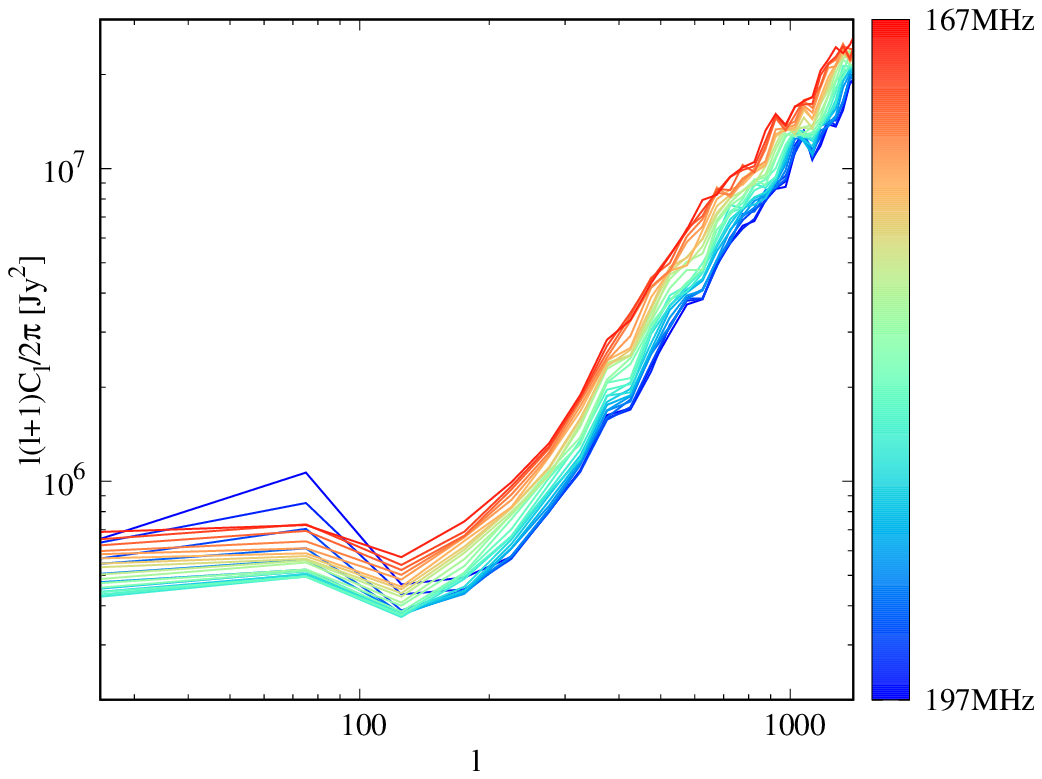}
\caption{The APS computed from visibility data with 1.28~MHz bandwidth. The redder line shows the low frequency data and the bluer shows high frequency. At small scales, the high frequency APS is smaller than the low frequency APS. However, the some of APS at high frequency enhanced at large scales. This is caused by the leakage of Galactic synchrotron emission from the side lobe. } 
\label{fig:APSfreq} 
\end{figure}

Fig.~\ref{fig:APSfreq} shows the APS calculated from visibility data with 1.28MHz bandwidth in units of $\rm Jy^2$. 
At small scale ($l>200$), the signal increases with decreasing frequency, and the APS at 167MHz is 1.53 times larger than the signal at 197MHz. 
This feature is consistent with the negative spectral index of point source foregrounds \citep[e.g., ][]{2017MNRAS.464.1146H}.
On the other hand, at large scales, the several APSs at high frequencies are stronger than that of low frequency data. 
As discussed above, the APS is dominated by the Galactic synchrotron emission with the negative spectral index at these scale, and therefore this feature is un-expected. This feature is caused from the pointing dependence of APS as shown in Fig.~\ref{fig:APSLST}.

\begin{figure}
\centering
\includegraphics[width=7.5cm]{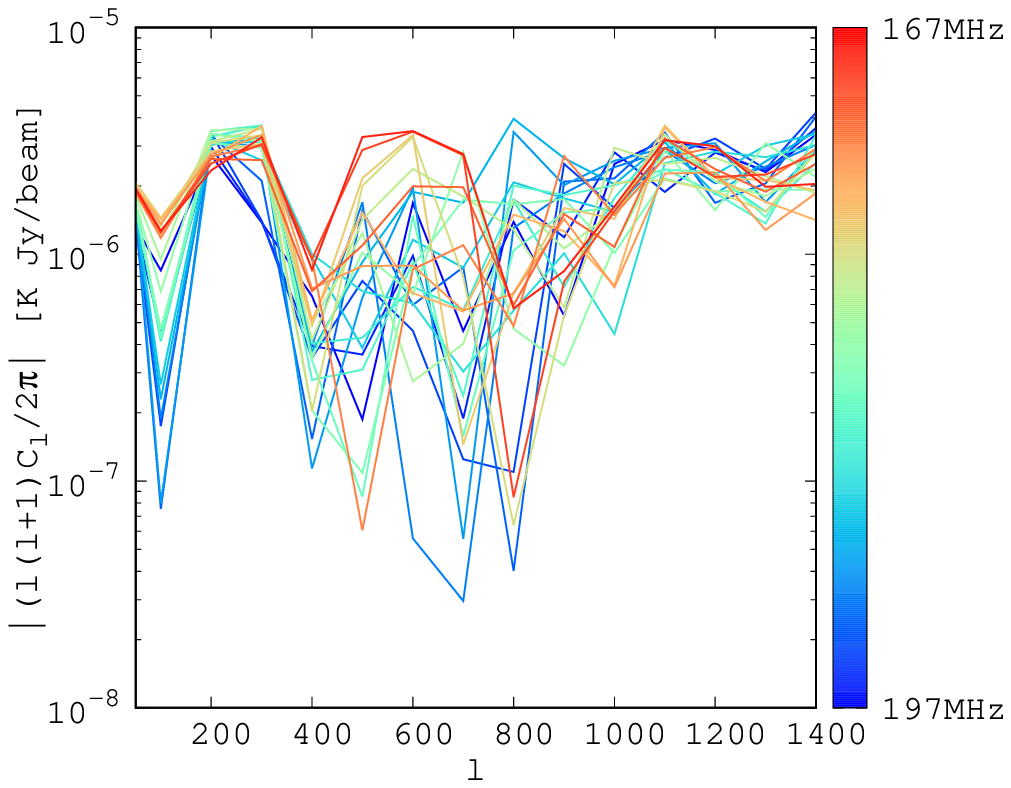}
\includegraphics[width=7.5cm]{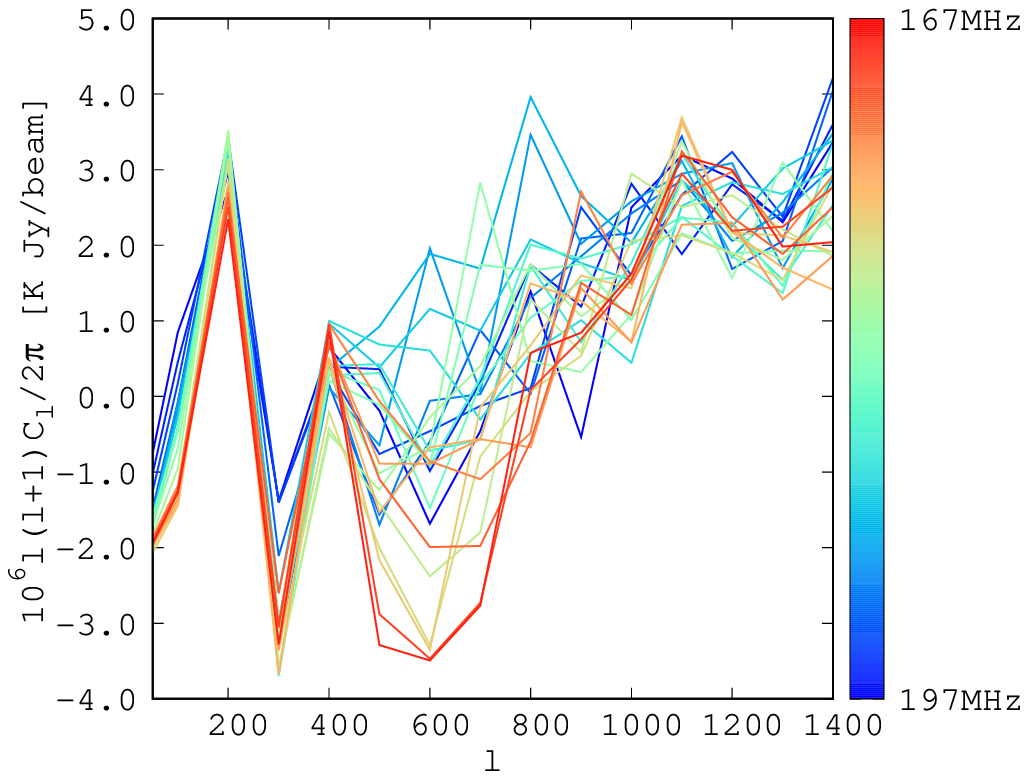}
\caption{{Same as Fig.~\ref{fig:APSfreq}, but for the CPS calculated from images. We show the absolute value in top panel, and the bottom panel shows the signal in linear scale. }}
\label{fig:CPSfreq} 
\end{figure}

{Frequency dependence of the CPS is shown in the Fig.~\ref{fig:CPSfreq}. No clear tendency can be seen in the behavior of the CPS. However, the CPS is dominated by foreground error, as indicated in the above analysis (e.g. Fig.~\ref{fig:IonoX} and Fig.~\ref{fig:CPSLST}). Thus, Fig.~\ref{fig:CPSfreq} indicates that the frequency dependence is weaker than the variance due to foreground errors. }{However, the CPS shows weak oscillation and frequency dependence at $l<300$. This result suggests that the CPS has only a few independent modes at large scales and the modes do not vary with frequency. Furthermore, at $l>1200$, the CPS is always positive, and this indicates unexpected systematics at small scales. }

\section{Results}\label{S:results}
{In this section, we discuss the APS and CPS from the MWA data and the Planck CMB map after 3 hours integration.}

\subsection{Angular power spectrum}

\begin{figure*}
\centering
\hspace{0cm}\includegraphics[width=7cm]{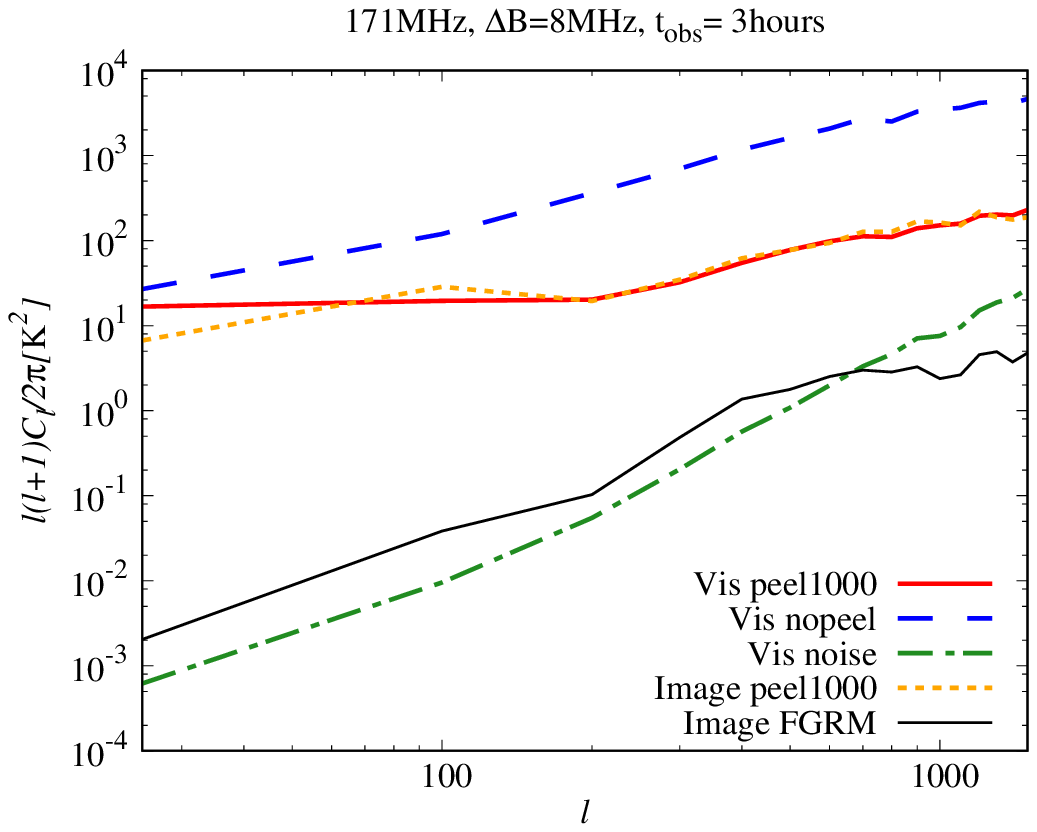}
\hspace{1cm}\includegraphics[width=7cm]{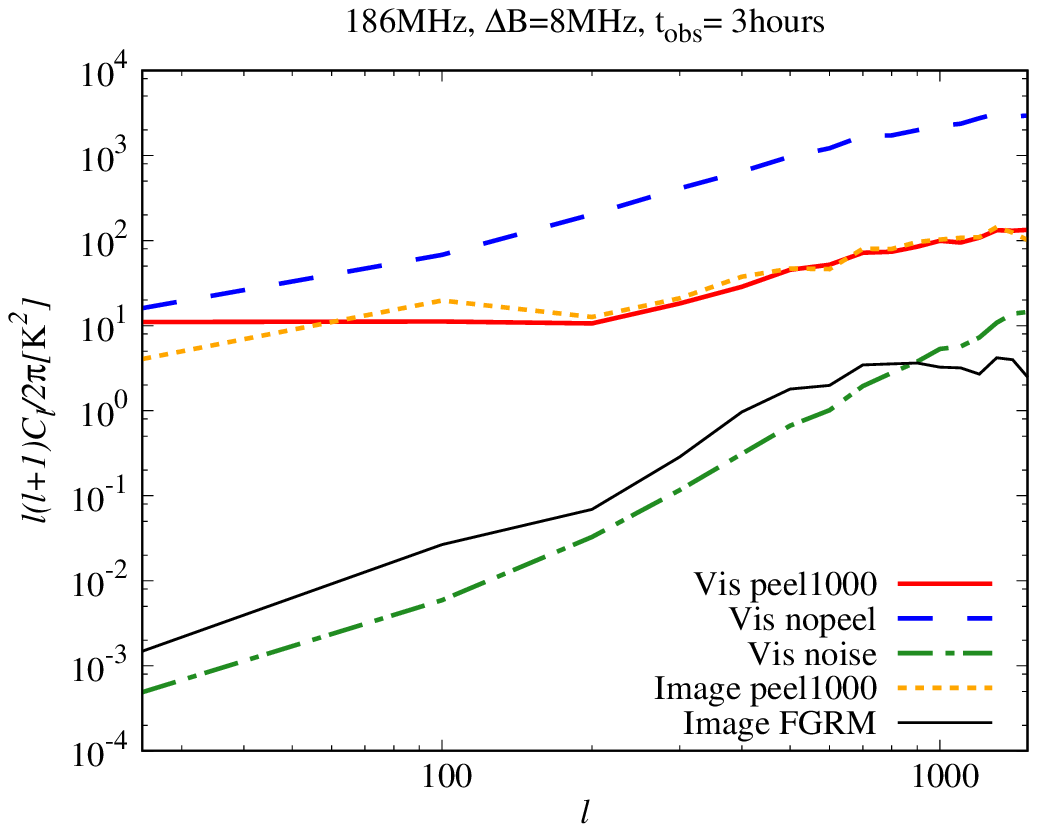}
\hspace{0cm}\includegraphics[width=7cm]{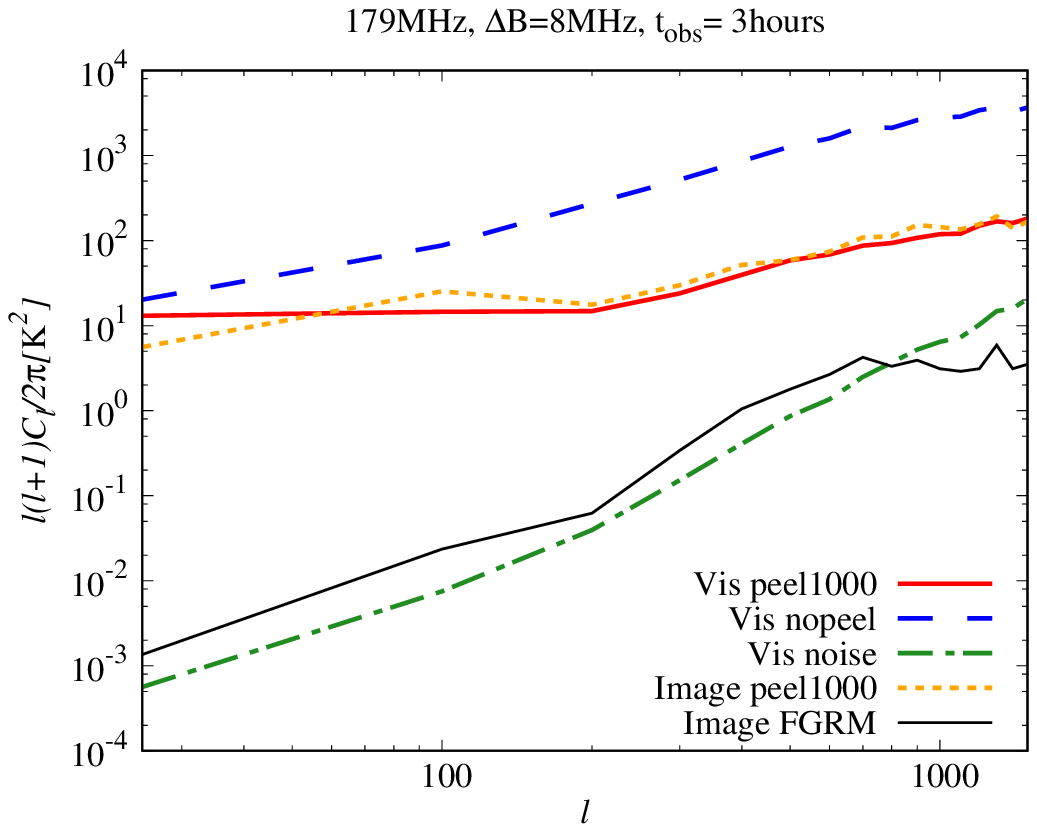}
\hspace{1cm}\includegraphics[width=7cm]{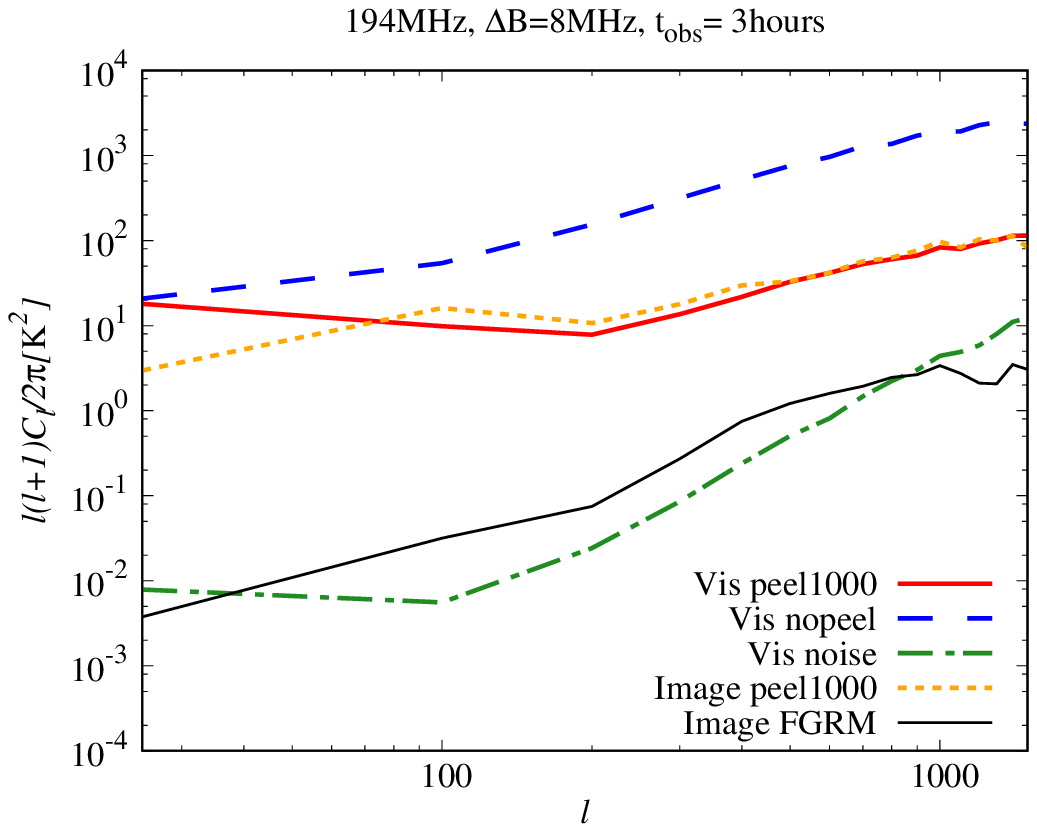}
\caption{Comparison of the APS. Thick solid line shows the APS of visibility with peeling of 1000 point sources. The dashed line is also the APS of visibility without peeling. The dot-dashed line is the noise level estimated from the visibility data. The APS of our main images is shown as dotted line. The thin solid line is calculated from the foreground removed images described in Sec.~\ref{S:FGRM}.  } 
\label{fig:APSimage} 
\end{figure*}

To understand the MWA data, we calculate the APS using different 5 data sets. The 5 data sets are: i) the visibility with no source peeling, ii) visibility with peeling of 1000 sources, iii) image with peeling of 1000 sources, iv) image with the polynomial foreground removal and peeling of 1000 sources and v) visibility of the thermal noise. Fig.~\ref{fig:APSimage} compares the resulting APS at $\nu=$ 167MHz, 176MHz, 186MHz and 194MHz. 

The APS using visibilities without peeling increases with $l$ and overwhelms the APS with peeling of 1000 sources at all scales. Thus, the apparent 1000 point sources dominate the signal. Meanwhile, the APS calculated from visibility with peeling shows a flat slope at large scales and rises at small scales. This different index can be understood as that the point source foregrounds dominate at small scales and the diffuse Galactic emission is powerful at large scales. Typically, the expected 21\,cm APS is less than $10^{-4}\rm [K^2]$ at $\ell = 100$ \citep{2007MNRAS.378..119D}, and therefore we have to reduce the foregrounds by at least 6 orders of magnitude in $\rm K^2$ to measure the 21\,cm-line signal. 

The noise APS of the MWA observation is sufficiently smaller than the measured signal dominated by the foreground. However, the noise is larger than the expected 21\,cm line signal. Thus, in addition to the precise foreground removal, we need a deep integration in time for the measurement of 21\,cm-line signal.
 
 {The APS calculated from image with peeling 1000 sources should be consistent with the APS of the visibility since these data have the calibration and the point source subtraction in common. Indeed, the Fig.~\ref{fig:APSimage} shows that the APS of image is fairly consistent with the APS of visibility. However, the APS of image is smaller than the APS of visibility at $\ell = 25$.} This could be because that the APS of visibility is enhanced by the primary beam shape of the MWA(e.g. Fig.~1 in \cite{2016ApJ...818..139T}) but the the weight of mask can be corrected in the APS of images by using Polspice. 

{By using the polynomial fitting, we can reduce the foreground contamination in the APS. If the foreground removal is accurate, the resulting APS should be consistent with the thermal noise contamination. However, the APS of the foreground removed image is larger than the noise level since our simple foreground removal is not perfect. Nevertheless, the APS of the foreground removed image is 3 orders of magnitude smaller than the APS with peeling of 1000 sources at $l=100$}. Furthermore, the  APS is reduced 1$\sim$2 orders of magnitude at small scales. This indicates that the polynomial fitting removes both the diffuse emission and the point sources. However, the remaining signal is larger than the expected signal. Thus, other accurate foreground removal is required to remove the foreground residuals. 

{It should be noted that the APS with the foreground removal is smaller than the noise level at small scales. This unexpected result could be caused from overfitting. At the small scales, the thermal noise is 10\% of total signal and then the thermal noise could be fitted by the polynomial function. } 

\begin{figure}
\centering
\includegraphics[width=7cm]{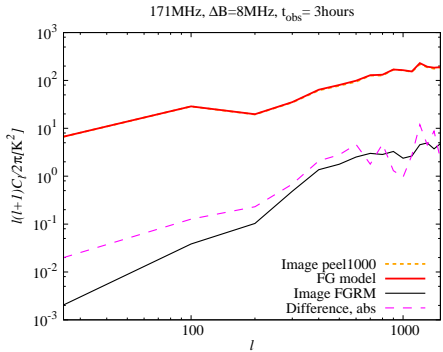}
\caption{Comparison of the APS of our main images (dotted line) and the APS of a foreground model obtained by the polynomial fitting (thick solid line) at 171~MHz.  The dashed line is the absolute value of the difference between the dotted and solid lines. } 
\label{fig:APSFGmodel} 
\end{figure}

We compute the APS of the foreground model obtained using the polynomial fitting. Fig.~\ref{fig:APSFGmodel} shows that the APS of the foreground model is consistent with the APS of the main image at 171~MHz. The amplitude of the difference is larger than the APS of the residual image at large scales. This is indicative of the variance due to the cross term of the foreground model and the residual.

\subsection{21\,cm-CMB Cross Power Spectrum}\label{S:Cross}

\begin{figure*}
\centering
\includegraphics[width=15cm]{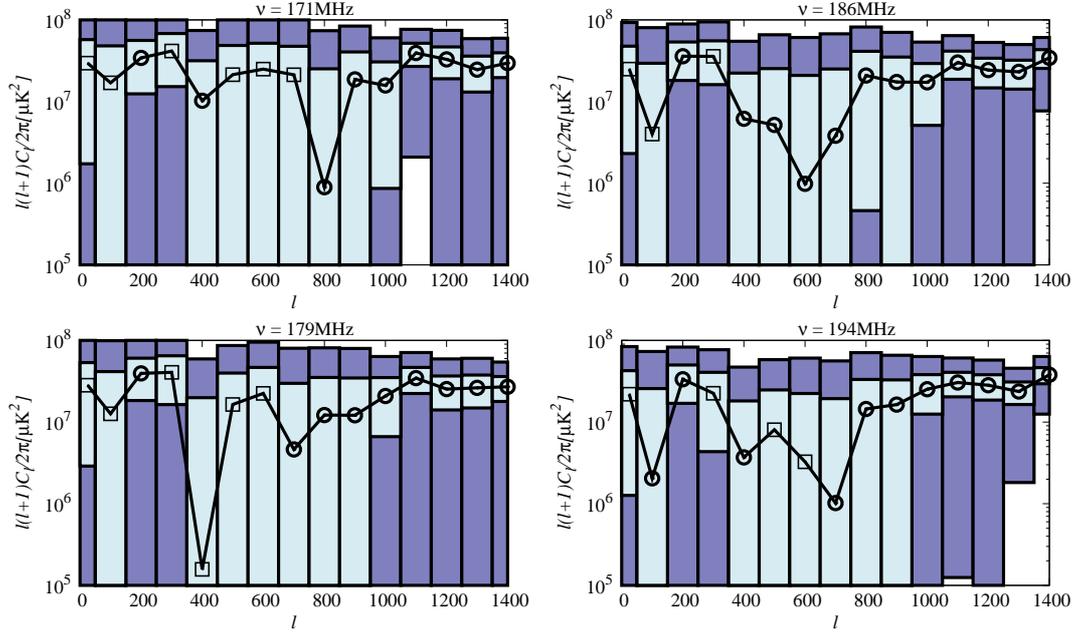}
\caption{CPS between the MWA observed image with point source subtraction and the Planck CMB image. Black solid line is the CPS, circles indicate positive values and squares indicate negative. Boxes are the expected 1 $\sigma$ and 3 $\sigma$ error, which are calculated using Eq.~\ref{eq:error1}. Where the boxes meet the $x$-axis, the signal is consistent with zero. At $l<1000$, there are no significant detections. At small scales, unexpected signals are detected with high significance and positive correlation. } 
\label{fig:cps1} 
\end{figure*}

Fig.~\ref{fig:cps1} shows the resulting CPS at each frequency without the polynomial foreground removal, with the 1 sigma and 3 sigma error estimated using Eq.~\ref{eq:error1}, where $\Delta l=100$. There is no-detection of any signal at $l<1000$ within a 3 sigma error. Circles and squares represent the negative and positive sign, and the change of sign is a result of large error dominated by the foregrounds. Also, this indicates that there is not strong systematics such as the correlation between the foregrounds of the 21\,cm observation and the foreground residual in the CMB data. The expected 21\,cm-CMB CPS has a peak at $l=100$, and we obtain $3\times10^7 \rm \mu K^2$ {as the 1$\sigma$ error} on the scale. Table.~\ref{table:upper} lists the measured signal and error at different scales and frequencies. 

The foreground emission is expected to be weak at high frequency due to the negative spectral index of synchrotron emission. As anticipated, the error and the amplitude of the measured signal are large at low frequency on large scales. We note that the signal is contaminated by large variance.

At $l>1000$, we find detection of an unexpected signal with 2$\sim$3 $\sigma$ significance, especially at 194~MHz. However, at these scales, the foreground residual of the CMB is small and therefore the correlation of foregrounds is unlikely to be the source of the detection. Thus, this detection suggests that we underestimate the error on the CPS since the Eq.~\ref{eq:error1} is valid only if the error follows a Gaussian distribution. 

\begin{figure}
\centering
\includegraphics[width=7cm]{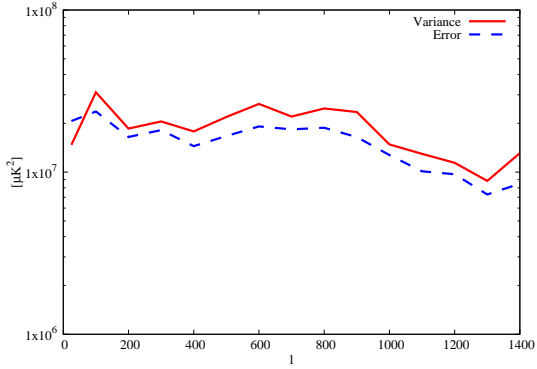}
\caption{Comparison of the error at 194~MHz. The solid line is the variance of the CPS and the dashed line is error estimated by Eq.~\ref{eq:error1}. For estimating the variance, we calculate the CPS between the MWA image and 100 random CMB maps.  } 
\label{fig:Errsamp} 
\end{figure}

In order to evaluate the error numerically, we attempt to calculate the CPS between the MWA image and 100 different CMB images. Fig.~\ref{fig:Errsamp} shows comparison between the variance of signals and the error estimated by the Eq.~\ref{eq:error1} at $194$ MHz. The variance is consistently larger than the expected error at $l>200$. This indicates an additional contamination to the error term, such as a non-Gaussian error and {unexpected systematics}. This error could {caused} from the point sources which dominates the APS at the scales. On the other hand, the variance is fairly consistent with the expected error at $l \approx 100$. This indicates that the systematics is small on large scales. Note that the variance is a conditional error since we fix the MWA image. Thus, we use the error of Eq.~\ref{eq:error1} for the analysis.

\begin{figure*}
\centering
\includegraphics[width=15cm]{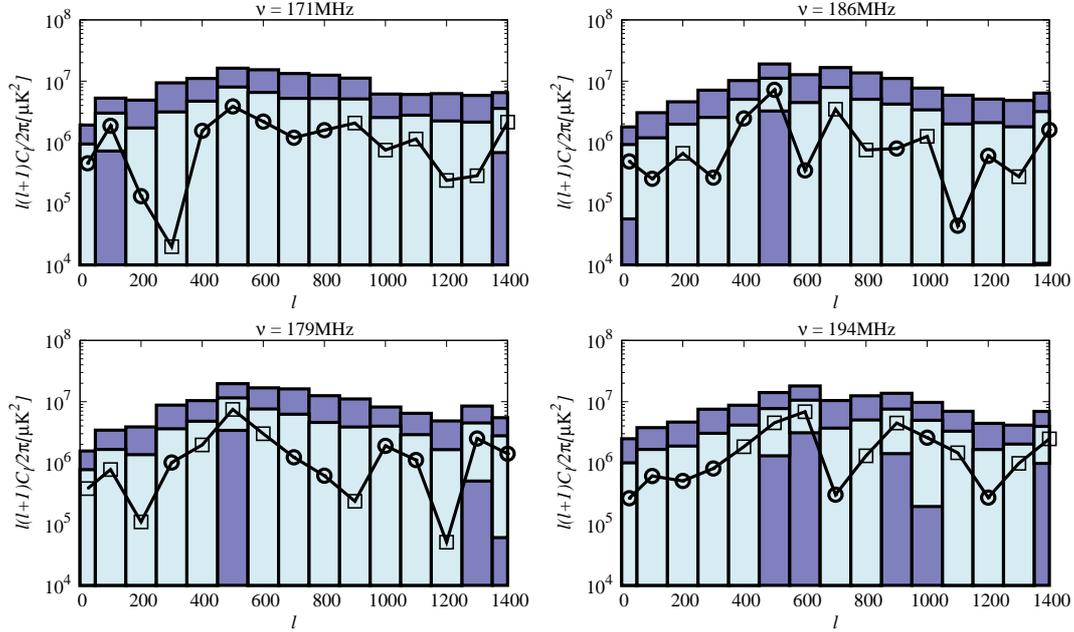}
\caption{Same as Fig.~\ref{fig:cps1}, but the foregrounds are removed by the polynomial fitting with $n$=7. There are no significant detections even at $l>1000$. } 
\label{fig:sum} 
\end{figure*}

The Fig.~\ref{fig:sum} shows the results with images using the polynomial foreground removal. At all scales, there is no significant detection and the error dominates the signal. However, {the CPS and the error} are reduced to 1 order of magnitude weaker than that of Fig.~\ref{fig:cps1}. We mention that our foreground removal method is simple and this method can provide critical signal loss and un-expected systematics especially at small scales ($l >1000$) as shown in the Fig.~\ref{fig:APSimage}. In Table.~\ref{table:upper2}, we summaries the measured values and the 1 $\sigma$ error. 

Same as Fig.~\ref{fig:Errsamp}, we compare the variance with error and find that the variance of the CPS with the polynomial foreground removal is consistent with the estimated error at all scales. This indicates that {unexpected systematic} error is likely due to the foregrounds.

\begin{table*}
  \centering
  \begin{tabular}{|l|c|c|c|c|} \hline
     & $l$=100 &$l$=200 & $l$=500 & $l$=1000 \\ \hline \hline
    171MHz & $-1.696\times 10^7\pm3.117\times 10^7$ & $3.423\times 10^7\pm2.181\times 10^7$ & $-2.135\times 10^7\pm2.731\times 10^7$ & $1.575\times 10^7\pm1.289\times 10^7$\\ \hline
    179MHz & $-1.262\times 10^7\pm2.885\times 10^7$ & $3.943\times 10^7\pm2.107\times 10^7$ & $-1.642\times 10^7\pm2.334\times 10^7$ & $2.083\times 10^7\pm1.420\times 10^7$\\ \hline
    186MHz & $-4.004\times 10^6\pm2.545\times 10^7$ & $3.589\times 10^7\pm1.777\times 10^7$ & $5.164\times 10^6\pm2.018\times 10^7$ & $1.720\times 10^7\pm1.209\times 10^7$\\ \hline
    194MHz & $2.044\times 10^6\pm2.367\times 10^7$ & $3.345\times 10^7\pm1.648\times 10^7$ & $-8.054\times 10^6\pm1.670\times 10^7$ & $2.526\times 10^7\pm1.275\times 10^7$\\ \hline
  \end{tabular}
  \caption{List of the cross power spectrum of 21\,cm-CMB and 1 $\sigma$ error at different scales and frequencies. Using the RTS, 1000 point sources have been removed. }
  \label{table:upper}
\end{table*}

\begin{table*}
  \centering
  \begin{tabular}{|l|c|c|c|c|} \hline
     & $l$=100 &$l$=200 & $l$=500 & $l$=1000 \\ \hline \hline
    171MHz & $1.864\times 10^6\pm1.141\times 10^6$ & $1.325\times 10^5\pm1.586\times 10^6$ & $3.875\times 10^6\pm4.134\times 10^6$ & $-7.488\times 10^5\pm1.800\times 10^6$\\ \hline
    179MHz & $-7.779\times 10^5\pm8.806\times 10^5$ & $-1.088\times 10^5\pm1.253\times 10^6$ & $-7.466\times 10^6\pm4.064\times 10^6$ & $1.893\times 10^6\pm2.089\times 10^6$\\ \hline
    186MHz & $2.546\times 10^5\pm9.343\times 10^5$ & $-6.634\times 10^5\pm1.313\times 10^6$ & $7.203\times 10^6\pm3.944\times 10^6$ & $-1.253\times 10^6\pm2.149\times 10^6$\\ \hline
    194MHz & $6.090\times 10^5\pm1.049\times 10^6$ & $5.070\times 10^5\pm1.377\times 10^6$ & $-4.504\times 10^6\pm3.198\times 10^6$ & $2.573\times 10^6\pm2.379\times 10^6$\\ \hline
  \end{tabular}
  \caption{Same as Table.~\ref{table:upper}, but we remove the foregrounds from the MWA data by the polynomial fitting method.}
  \label{table:upper2}
\end{table*}


\section{Forecast}\label{S:discuss}

We discuss the feasibility of measuring the 21\,cm-CMB CPS based on the results. 
Previous works predict that the signal has peak at $l=100$. {For example, \cite{2006ApJ...647..840A} showed the CPS between 21\,cm signal and Doppler anisotropy is -200 $\rm \mu K^2$ at $z=7.5$. However, the correlation between 21\,cm signal and CMB anisotropy is caused from not only the Doppler effect but also Ostriker-Vishniac (OV) effects and patchy reionization (e.g. \cite{2016ApJ...824..118A}). {We mention that the fluctuation due to the Doppler effect dominates the kSZ anisotropy at large scale $l<100$, and the patchy and the OV anisotropy contribute at $l>200$.} Furthermore, the amplitude strongly depends on the reionization model, and therefore  we assume the 21\,cm-CMB CPS is $10^3$ $\rm \mu K^2$ at the scale as an optimistic prediction. }

The error is $3\times10^7$ $\rm \mu K^2$ for the data with peeling of 1000 point sources, and therefore the expected signal to noise ratio (SNR) of the result is $3.3\times 10^{-5}$. The foreground removal of the polynomial fitting reduces the error and the SNR is $10^{-3}$. 

Using the noise APS and Eq.~\ref{eq:error1}, we can estimate the SNR under the assumption that we have a perfect foreground removal.  
At $l=100$, the noise APS and the CMB APS are $10^{10}\rm [\mu K^2] $ and $3000\rm [\mu K^2]$. Thus, the error term of the thermal noise and the CMB APS is $4\times 10^5 \rm [\mu K^2]$. Although the noise contribution is smaller than the foreground contamination, the expected signal is 400 times smaller than the error caused by the thermal noise. 

\begin{figure*}
\centering
\includegraphics[width=15.5cm]{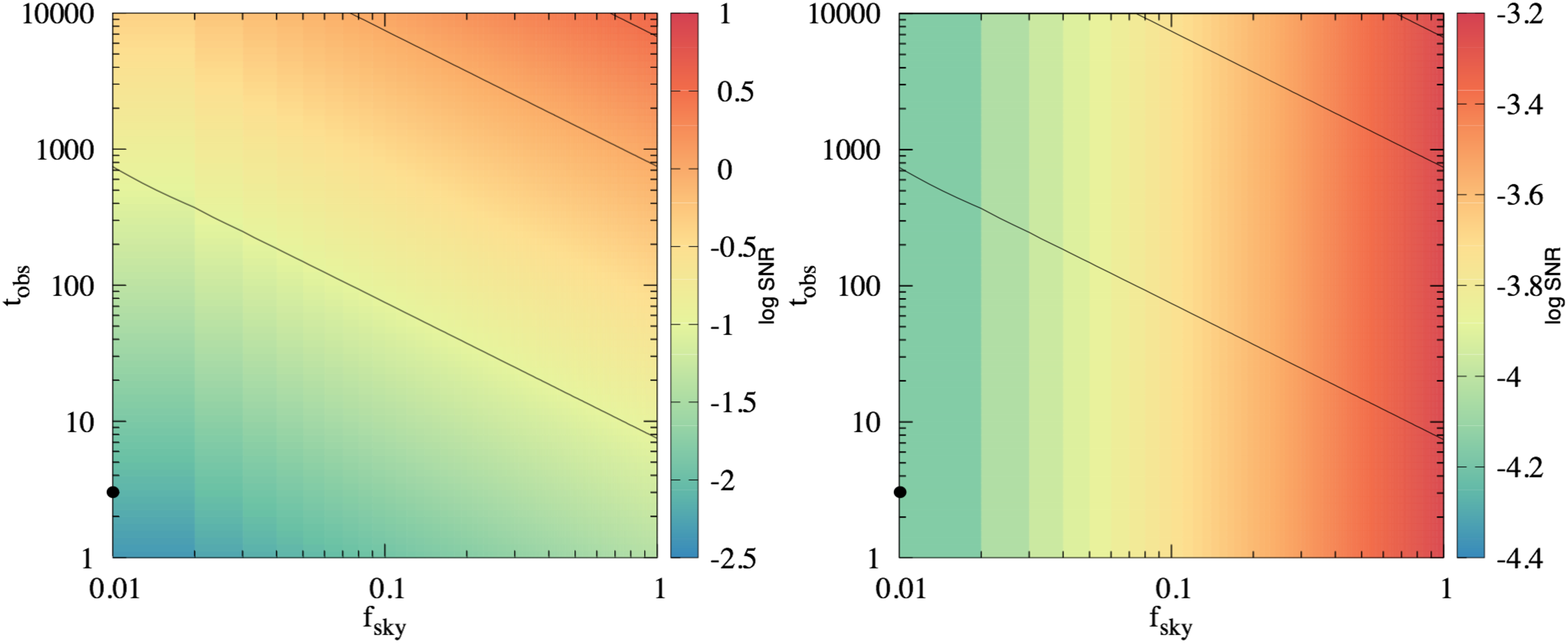}
\caption{Expected signal to noise ratio at $l=100$. The foreground is ignored (included) in left (right) panel. The three solid lines indicates that the SNR, without foreground contamination, are 0.1, 1 and 3 from bottom to top. We assume the 21\,cm-CMB CPS is $10^3\rm \mu K^2$. The error is estimated from Eq~\ref{eq:error1} with $\Delta l = 100$. Black point indicates SNR of this work. In the right panel, the value of $\log \rm SNR$ = -4 indicates that one has to remove the 99.99\% of foregrounds in units of Kelvin to detect the signal. } 
\label{fig:SNR} 
\end{figure*}

In order to increase the SNR, precise foreground removal and deep observations are required. Also, a large field of view can reduce the error of the 21\,cm-CMB CPS, which is proportional to $f_{\rm sky}^{-0.5}$. Figs.~\ref{fig:SNR} show the contour of the expected SNR at $l=100$ in the observation time and the sky fraction plane. The SNR at $f_{\rm sky} = 0.01$ and $t_{\rm obs}=3\rm hrs$ corresponds to the SNR of this work.
In the left panel, we assume the foreground is perfectly removed. To achieve the SNR$\approx$1, the MWA needs 2000 hrs of observation for each MWA field of view and 50 \% of sky fraction, but such enormous survey is not realistic. Thus, more sensitive instruments are necessary for the measurement. Fortunately, the MWA Phase II has additional tiles which form two hexagonal arrays and high sensitivity at large scale. 

The right panel of Fig.~\ref{fig:SNR} shows the SNR including the foreground contamination except removed apparent 1000 point sources. 
{Because the error is proportional to $\sqrt{C_l^{\rm FG}}$, we need a 99.95\% reduction of the foregrounds in units of Kelvin to achieve the SNR$\sim$1 with $f_{\rm sky}=0.5$.} 
Large $f_{\rm sky}$ can reduce the required precision of the foreground removal. Thus, instruments that are capable of observing with the large field of view, have advantage. For example, the instantaneous field of view of the MWA is 100 times larger than the SKA1\_LOW, and the required accuracy of foreground removal is 10 times lower than that of the SKA1\_LOW. In practice, the drift scan strategy with high sensitivity can compensate for the disadvantage. 

The 21\,cm APS is also a source of the error term. As discussed in \cite{2008MNRAS.384..291A}, the severity of the error term depends on the correlation coefficient. Ignoring other error terms, the SNR is approximately given by ${\rm SNR}^2 \approx r^2_{\rm 21,D}\Delta l (2l+1)f_{\rm sky} C_l^{\rm D}/C_l^{\rm CMB}$, where  $r_{\rm 21,D}$ is cross correlation coefficient between 21\,cm line signal and the Doppler anisotropy and $C_l^{\rm D}$ is the APS of the Doppler contribution. Let us assume $C_l^{\rm D}/C_l^{\rm CMB}\approx 1/3000 $, $\Delta l = 200$ at $l=100$ and the 21\,cm line and the Doppler anisotropy are correlate perfectly, $r_{\rm 21,D}=1$. Although the assumption is optimistic, the $f_{\rm sky}$ must be larger than 0.7 to achieve the detection with high significance (SNR>3). This pessimistic situation is improved at high-$z$ for the early reionization model as the fraction between $C_l^{\rm D}$ and $C_l^{\rm cmb}$ could increase \citep{2008MNRAS.384..291A}.

The expected 21\,cm-CMB CPS tends to be powerful at higher redshift. For example, the CPS can be $10^4\rm \mu K^2$ at $z\sim10$ for the early reionization model (e.g. \cite{2016ApJ...824..118A}). If the spectral index of the foreground and thermal noise are $-2.6$, the error at $z=10$ is 2 times larger than the error at 171~MHz. Thus, the required survey area, the observation time and the precision of foreground removal are small. The analysis using lower frequency data is our future work, and it could be used to constrain the extreme reionization model.

\section{Summary}\label{S:summary}

{In this work, we have presented a first attempt to calculate the cross power spectrum (CPS) between 21\,cm line signal and the CMB using the 3 hours of MWA data and the CMB temperature map observed by the Planck. For reducing the error, dominated by the foregrounds, we have subtracted 1000 point sources with the RTS and have removed spectrally smooth foregrounds by a polynomial fitting across frequency channels. }

{
The 21\,cm observation with the interferometers is complicated due to Earth's ionosphere, observation pointing, foreground and frequency dependence. Thus, we have investigated the systematic contamination to the CPS. We could not find any correlation of results with level of ionospheric activity, or a dependency on frequency, but the CPS is dominated by the foreground error depending on the group of pointings. However, we have used only 3 hours of data for the analysis. In future work, a larger amount of data is required.
}

{
By computing the angular power spectrum (APS) using the image and visibility, we found that the Galactic foreground and point sources dominate the signal at large, $l\sim100$, and small scales, $l>200$. Although the simple polynomial fitting can reduce about 90\% of foreground, the residual was larger than the noise level.  
}

{
The CPS can statistically reduce the foreground contamination. However, the extremely bright foregrounds provided a large error on the CPS, even if we removed the foregrounds. Thus, we could not detect the 21\,cm-CMB signal and obtained at 68\% confidence interval of $\Delta C_l = 3.1\times 10^7$ at $l=100$ and at 171~MHz without the polynomial foreground removal. Furthermore the variance of the CPS is large than the expected error due to unexpected systematics at small scales. We also have found that the measurement of CPS requires at least 50\% of sky fraction, 2000 hours of observation time for the MWA field of view and removal of 99.95\% foreground in Kelvin. 
}

\section*{Acknowledgements}
The author thanks to J. L. B. Line for useful comments, {C. H. Jordan for providing the list of ionosphere activity} and all MWA EoR members for discussion at initial stage of this work. 
This work was supported by resources awarded under Astronomy Australia Ltds merit allocation scheme on the gSTAR national facility at Swinburne University of Technology. gSTAR is funded by Swinburne and the Australian Governments Education Investment Fund. This work is supported by JSPS KAKENHI Grant Numbers, JP16J01585 (S.Y.), JP15H05890 (K.I.), JP16H01543 (K.I.), JP26610048 (K.T.), JP15H05896 (K.T.), JP16H05999 (K.T.), JP17H01110 (I.K.,  K.T. and H.T.), JP15K17646 (H.T.), and Bilateral Joint Research Projects of JSPS (K.T.). The Centre for All Sky Astrophysics in 3D (ASTRO 3D) is an Australian Research Council Centre of Excellence, funded by grant CE170100013. CMT is supported under the Australian Research Council's Discovery Early Career Researcher funding scheme (project number DE140100316). The Centre for All-Sky Astrophysics (CAASTRO) is an Australian Research Council Centre of Excellence, funded by grant CE110001020.

\bibliographystyle{mnras}

\appendix

\section{Weight correction}

The image used in this work are provided with the natural weight to increase the sensitivity at large scales. Thus, the APS and CPS calculated from images is affected by the weight caused from uv-coverage. The output image can be approximately described as,
\begin{eqnarray}
I_{i} ({\bf{l}}) = \int U_i({\bf{u}})V({\bf{u}}) \exp(2\pi i {\bf{l}}\cdot{\bf{u}}) d^2u,
\end{eqnarray}
where $U$ is weight due to the uv-samplings and we ignore the w-term for simplicity. 
The main images are obtained as the average of different images, which is written as
\begin{eqnarray}
I_{\rm sum} ({\bf{l}}) &=& \frac{1}{N}\sum_{i} I_{i}\nonumber\\
&=& \frac{1}{N}\sum_{i} \int U_i({\bf{u}})V({\bf{u}}) \exp(2\pi i{\bf{l}}\cdot{\bf{u}}) d^2u\nonumber\\
&=& \int \tilde{U}({\bf{u}})V({\bf{u}}) \exp(2\pi i{\bf{l}}\cdot{\bf{u}}) d^2u,
\end{eqnarray}
where 
\begin{eqnarray}
\tilde{U} ({\bf{u}}) = \frac{1}{N} \sum_{i} U_i({\bf{u}}).
\end{eqnarray}
Furthermore, we assume the $V({\bf{u}})$ is consistent with all observations. 
Under the flat sky approximation, the integrated visibility is
\begin{eqnarray}
V_{\rm sum} = \int I_{\rm sum} \exp(-2\pi i {\bf{l}}\cdot{\bf{u}}) d^2l 
=\tilde{U} V,
\end{eqnarray}
and the power spectrum is given by
\begin{eqnarray}
C_{l,w}({\bf{u}}) \propto |V_{\rm sum}^2| = \tilde{U}^2 V^2.
\end{eqnarray}
By averaging over the $uv$-cells contributing to that $l$-mode ($l$ = 2$\pi$\,|${\bf{u}}$|), the APS is
\begin{eqnarray}
C_{l,w}(|{\bf{u}}|) \propto \frac{1}{m}{\displaystyle  \sum^{m}_{|{\bf{u}}|}} \tilde{U}^2 V^2. 
\end{eqnarray}
Finaly the correction for the APS can be written as
\begin{eqnarray}
C_l(|{\bf{u}}|) \propto  \frac{C_{l,w}(|{\bf{u}}|)}{\frac{1}{m}{\displaystyle \sum^{m}_{|{\bf{u}}|} }\tilde{U}^2}. 
\end{eqnarray}
Similarly, the correction for the CPS is given as,
\begin{eqnarray}
C_l^{X}(|{\bf{u}}|) \propto  \frac{C^{X}_{l,w}(|{\bf{u}}|)}{\frac{1}{m}{\displaystyle \sum^{m}_{|{\bf{u}}|} }\tilde{U}}.
\end{eqnarray}
Note that the $U$ is not squared since the CMB image has no effect from the uv samplings. 
To consider the error propagation, the error of the observed CPS is estimated as, 
\begin{eqnarray}
\sigma^2_{\rm X} = \frac{1}{(2l +1)f_{\rm sky} \Delta l} C_{l,w}^{\rm MWA}C_l^{\rm CMB}\left({\frac{1}{m}{\displaystyle \sum^{m}_{|{\bf{u}}|} }\tilde{U}}\right)^{-2}
\end{eqnarray}
, where $C_{l,w}^{\rm MWA}$ is the APS of weighted image. 
Although our correction of the weight is simple, the image APS can reproduce the APS computed from visibility.

\label{lastpage}
\end{document}